\documentclass[apj]{emulateapj_rtx4}

\usepackage{mathptmx}
\usepackage{epsfig,natbib}
\slugcomment{\rm To Appear in the {\sl Astrophysical Journal Supplements}}
\shortauthors{Eracleous, Hwang, \& Flohic}
\shorttitle{SEDs of Weak AGNs in LINERs}

\def\aj{\rm{AJ}}                   
\def\araa{\rm{ARA\&A}}             
\def\apj{\rm {ApJ}}                
\def\apjl{\rm{ApJ}}                
\def\apjs{\rm{ApJS}}               
\def\aap{\rm{A\&A}}                
\def\mnras{\rm{MNRAS}}             

\def\cd{c$\!\!\!\hskip 0.75pt$\raise 0.2pt \hbox{\symbol{24}}}

\def\Msol{\ifmmode{\rm M}_{\mathord\odot}\else M$_{\mathord\odot}$\fi}
\def\Mbh{\ifmmode{M_{\rm BH}}\else{$M_{\rm BH}$}\fi}
\def\REdd{\ifmmode{{\cal R}_{\rm Edd}}\else{${\cal R}_{\rm Edd}$}\fi}
\def\nLn{\ifmmode{\nu L_{\nu}}\else{$\nu L_{\nu}$}\fi}

\def\ls{\lower 2pt \hbox{$\;\scriptscriptstyle \buildrel<\over\sim\;$}} 
\def\gs{\lower 2pt \hbox{$\;\scriptscriptstyle \buildrel>\over\sim\;$}}

\def\sed{SED}
\def\seds{SEDs}

\def\kms{\ifmmode{{\rm km~s^{-1}}}\else{km~s$^{-1}$}\fi}
\def\ergs{\ifmmode{{\rm erg~s^{-1}}}\else{erg~s$^{-1}$}\fi}
\def\m#1{\ifmmode{^{-#1}}\else{$^{-#1}$}\fi}
\def\asec{\ifmmode{^{\prime\prime}}\else{$^{\prime\prime}$}\fi}
\def\asecb{\ifmmode{^{\prime\prime\!\!\!}}\else{$^{\prime\prime\!\!\!}$}\fi}
\def\asecp{\ifmmode{^{\prime\prime\!\!\!}.}\else{$^{\prime\prime\!\!\!}$.}\fi}
\def\deg{\ifmmode{^{\circ}}\else{$^{\circ}$}\fi}
\def\degp{\ifmmode{^{\circ\!\!\!}.}\else{$^{\circ\!\!\!}$.}\fi}
\def\aox{\ifmmode{\alpha_{\rm ox}}\else{$\alpha_{\rm ox}$}\fi}
\def\zbol{\ifmmode{\kappa_{\rm 2-10\; keV}}\else{$\kappa_{\rm 2-10\; keV}$}\fi}

\def\ten#1{\ifmmode{\times 10^{#1}}\else{$\times 10^{#1}$}\fi}
\def\tten#1{\ifmmode{\times 10^{#1}}\else{$\times 10^{#1}$}\fi}
\def\nten#1#2{\ifmmode{#1\times 10^{#2}}\else{$#1\times 10^{#2}$}\fi}

\newcounter{species}
\def\ion#1#2{\setcounter{species}{#2}#1$\;${\sc\roman{species}}\relax}

\def\flion#1#2#3{[{\setcounter{species}{#2}#1$\;${\sc\roman{species}}]$\;\lambda${#3}}\relax}

\def\a{$\alpha$}
\def\b{$\beta$}

\def\xmm{{\it XMM-Newton}}
\def\chandra{{\it Chandra}}

\def\hst{{\it HST}}

\def\ptwo{paper II}

\begin{document}

\title{Spectral Energy Distributions of Weak Active Galactic Nuclei
  Associated With Low-Ionization Nuclear Emission Regions}

\author{Michael Eracleous\altaffilmark{1,2,3}, 
        Jason A. Hwang\altaffilmark{1,3}, \& 
        H\'el\`ene M. L. G. Flohic\altaffilmark{1,4}}

\altaffiltext{1}{Department of Astronomy and Astrophysics, The Pennsylvania
State University, 525 Davey Lab, University Park, PA 16802} 

\altaffiltext{2}{Center for Gravitational Wave Physics, The Pennsylvania
State University, 104 Davey Lab, University Park, PA 16802}

\altaffiltext{3}{Department of Physics \& Astronomy, Northwestern
University, 2131 Tech Drive, Evanston, IL 60208}

\altaffiltext{4}{Department of Physics and Astronomy, University of
  California, 4129 Frederick Reines Hall, Irvine, CA 92697}

\begin{abstract}
We present a compilation of spectral energy distributions (\seds) of
35 weak active galactic nuclei (AGNs) in low-ionization nuclear
emission region (LINERs) using recent data from the published
literature. We make use of previously published compilations of data,
after complementing and extending them with more recent data. The main
improvement in the recent data is afforded by high-spatial resolution
observations with the {\it Chandra X-Ray Observatory} and high-spatial
resolution radio observations utilizing a number of facilities. In
addition, a considerable number of objects have been observed with the
{\it Hubble Space Telescope} in the near-IR through near-UV bands
since the earlier compilations were published. The data include upper
limits resulting from either non-detections or observations at low
spatial resolution that do not isolate the AGN.  For the sake of
completeness, we also compute and present a number of quantities from
the data, such as optical-to-X-ray spectral indices (\aox), bolometric
corrections, bolometric luminosities, Eddington ratios, and the
average \sed. We anticipate that these data will be useful for a
number of applications. In a companion paper, we use a subset of these
data ourselves to assess the energy budgets of LINERs.
\end{abstract}

\keywords{galaxies: nuclei --- galaxies: active --- X-rays: galaxies}

\section{Introduction}\label{S:intro}

Low-ionization nuclear emission regions (LINERs) were identified as a
class by \citet{heckman80} based on the relative intensities of their
oxygen emission lines. Their defining properties are:
\flion{O}{2}{3727}$\;/\;$\flion{O}{3}{5007}$\;> 1$ and
\flion{O}{1}{6300}$\;/\;$\flion{O}{3}{5007}$\;> 1/3$.  Optical
spectroscopic surveys \citep*[e.g.,][]{ho97a} have shown them to be
very common, occurring in approximately 50\% of the nuclei of nearby
galaxies. Suggestions for the ultimate power source of the emission
lines include (a) a weak active galactic nucleus \citep[an AGN
harboring an accreting, supermassive black hole; e.g.][]{halpern83,
ferland83}, (b) hot stars \citep[either young or old,
e.g.][]{terlevich85,filippenko92,shields92,barth00,binette94}, and (c)
shocks \citep[e.g.][and references therein]{heckman80, dopita96}.
Recent radio, UV, and X-ray surveys at high spatial resolution, mid-IR 
spectroscopy, and 
variability studies have uncovered weak AGNs in the majority of LINERs
studied so far, suggesting that they make up a large (perhaps the
largest) subset of all AGNs \citep*[e.g.][]{filho04, nagar05, barth98,
maoz05, ho01, terashima03, dudik05, flohic06, gonzalez06, dudik09}

The ubiquity of LINERs suggests that they are an important component
of the nuclei of galaxies in the local universe. Moreover, they trace
the population of AGNs at the lowest luminosities.  The data available
from the most modern observing facilities allow us to address a number
of outstanding questions related to LINERs and their central engines,
such as whether the weak AGNs are responsible for powering the
emission-line regions by photoionization, whether or not models for
low-radiative efficiency accretion flows provide a good description of
the properties of these weak AGNs, and what is the role of feedback of
the weak AGN on the central region of its host galaxy \citep[see the
  discussion of these and other related issues in][]{ho08}. A good
example of such an application is the work of \citet{maoz07}, who
independently collected measurements at very specific frequencies for
a subset of the objects in our sample.

Motivated by the above questions we set out to collect published data
from the recent literature to define the spectral energy distributions
(\seds) of weak AGNs in LINERs at photon energies above 1~Ry.  Our
primary interest was to assess the energy budgets of LINERs and
investigate whether the AGN is powerful enough to power the the
observed emission lines by photoionization.  In the process we
constructed the full \seds\ (from radio to X-ray energies), thus
extending and updating the pioneering work of \citet{ho99}.  In this
paper we present the data and a number of quantities derived from them
without any scientific interpretation. We anticipate that this
compilation of data will be useful for a number of additional
applications, including those mentioned in the previous paragraph. In
a followup paper we use a subset of the data presented here to address
the specific question regarding the energy budget of LINERs
\citep[][hereafter \ptwo]{eracleous09}.

In \S\ref{S:sample} we present the sample of galaxies and their basic
properties. In \S\ref{S:sed} we present the \seds, discuss extinction
corrections, and comment on a number of objects that warrant special
attention. Finally, in \S\ref{S:derived} we present a number of
quantities derived from the data, namely the optical-to-X-ray spectral
indices (\aox), the bolometric luminosities, Eddington ratios, and the
average \sed. We provide details of how we selected the data and how
we converted the measured quantities to ``monochromatic''
luminosities. We emphasize caveats and sources of uncertainty in our
methods. We also catalog the original data (with appropriate
references to the source of the information) so that one can reproduce
our procedures with modifications, if desired.

\section{Targets and Their Properties}\label{S:sample}

We constructed a sample of objects classified as LINERs or transition
objects by \citet{ho97b}\footnote{Hereafter, we refer to all objects
collectively as ``LINERs.'' In \S\ref{S:types} we re-examine their
classifications based on more recently developed criteria and
conclude that they can all be regarded as LINERs.} based on the
following considerations.  All objects were required to have measured
optical emission line luminosities and X-ray observations with
\chandra\ that yielded a measurement of the AGN X-ray spectrum or an
upper limit to its X-ray flux. The high spatial resolution of
\chandra\ allows us to spatially separate the AGN emission from most
of the diffuse, thermal X-ray emission.  We showed preference to
objects observed in the X-ray band with long exposure times i.e.,
objects whose X-ray spectra have high signal-to-noise ratio (S/N). In
such cases, a model fit to the AGN spectrum can give a direct estimate
of the total extinction toward the AGN continuum source. Moreover, any
soft, unresolved, circumnuclear X-ray emission can be separated
spectrally.  A significant subset of the sample objects were imaged in
the UV with the \hst\ and an overlapping subset had well sampled
\seds\ in the entire range from the radio to X-ray bands. We took
advantage of these overlapping subsets in order to determine upper
limits to the UV fluxes when they were not measured directly and to
determine bolometric luminosities from the data, as we detail in
\S\ref{S:sed}.

We relied heavily on previously published compilations of data
\citep[e.g.,][]{ho99,ptak04} and on papers presenting large
collections of measurements in the radio, UV, or X-ray bands
\citep[e.g.,][]{nagar05,doi05b,maoz05,flohic06}. We note that the work
of \citet{maoz05} is particularly important because it uses
variability in the UV continuum to check whether the nuclear source in
a LINER is an AGN.  The previously published compilations were updated
and supplemented with data that became available later. Because of
possible contamination of the AGN emission by circumnuclear emission,
we adopted measurements from high-spatial resolution observations
which isolate the AGN. In the near-IR, optical, and near-UV bands,
when given a choice we adopted measurements from images rather than
from spectra.  The high spatial resolution of \chandra\ and the \hst\
were optimal \citep[see example images in][]{eracleous02,flohic06}.
In the radio band, possible confusion from circumnuclear emission (or
in some cases jets) is also possible, therefore we relied on
high-resolution VLA or VLBA/VLBI observations.  For the sake of
completeness, we also include measurements made at low spatial
resolution, but we take these as upper limits to the flux of the AGN.
Included in the sample are
  13 LINERs with actual X-ray and UV (2500~\AA) measurements,
  11 LINERs with actual X-ray measurements but no UV measurements,
   6 LINERs with actual UV (2500~\AA) measurements and X-ray upper limits, and
   5 LINERs with X-ray upper limits only.

\begin{figure}
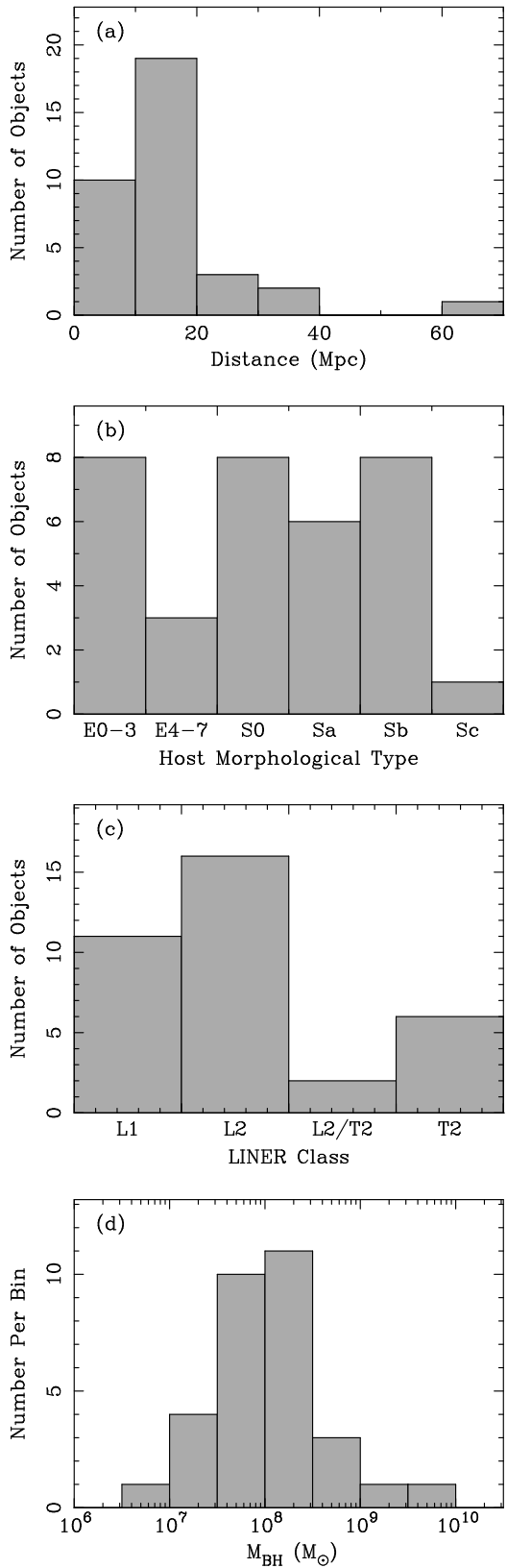

\centerline{\includegraphics[scale=0.4,angle=0]{f1a.eps}}  
\bigskip
\centerline{\includegraphics[scale=0.4,angle=0]{f1b.eps}} 
\bigskip
\centerline{\includegraphics[scale=0.4,angle=0]{f1c.eps}}  
\bigskip
\centerline{\includegraphics[scale=0.4,angle=0]{f1d.eps}} 
\caption{Distribution of the basic properties of the host galaxies of
the sample LINERs. The black hole masses in panel (d) were derived
from the stellar velocity dispersions, using
equation~(\ref{Q:bhmass}). NGC~266, NGC~3507, and NGC~4438 are not
included in the last histogram because their black hole masses are not
available. NGC~404 is also not plotted in the last histogram because
it is out of range, with $\log(M/{\rm M}_{\odot})=5.3$.
\label{F:hosts}}
\end{figure}

The galaxies of the resulting sample are listed in Table~\ref{T:hosts}
along with their morphological types and
distances. Table~\ref{T:nuclei} lists Galactic reddening, central
stellar velocity dispersion and inferred black hole mass, and LINER
type for the nuclei of the sample galaxies. The Galactic reddening was
taken from \citet{schlegel98}.  In Figure~\ref{F:hosts} we show the
distributions of the distances reported by \cite{tully88}, the
morphological types, the LINER types according to \cite{ho97b} of our
sample galaxies, and the black hole masses derived from the stellar
velocity dispersion. We describe and discuss the data below.

\begin{deluxetable}{llcccc}
\tabletypesize{\scriptsize}
\tablewidth{3.35in}
\tablecaption{Sample of Galaxies and Their Basic Properties\label{T:hosts}}
\tablehead{
\colhead{} &
\colhead{} &
\multicolumn{4}{c}{Distance (Mpc)} \\
\colhead{} &
\colhead{Hubble} &
\multicolumn{4}{c}{\hrulefill} \\
\colhead{Galaxy} &
\colhead{Type\tablenotemark{a}} &
\colhead{Tully\tablenotemark{b}} &
\colhead{SBF\tablenotemark{c}} &
\colhead{PNLF\tablenotemark{d}} &
\colhead{Other\tablenotemark{e}} \\
\colhead{(1)} &
\colhead{(2)} &
\colhead{(3)} &
\colhead{(4)} &
\colhead{(5)} &
\colhead{(6)}  
}
\startdata
  NGC 0266                 & SB(rs)ab   & 62.4 &      &      &        \\
  NGC 0404                 & SA(s)0     & ~2.4 & ~3.0 &      & ~3.1   \\
  NGC 1097                 & SB(rl)b    & 14.5 &      &      &        \\
  NGC 1553                 & SA(rl)0    & 13.4 & 17.2 &      &        \\
  NGC 2681                 & SBA(rs)0/a & 13.3 & 16.0 &      &        \\
  NGC 3031 (M81)           & SA(s)ab    & ~3.6 & ~3.6 &      & ~3.6   \\
  NGC 3169                 & SA(s)a     & 19.7 &      &      &        \\
  NGC 3226                 & E2         & 23.4 & 21.9 &      &        \\
  NGC 3379 (M105)          & E1         & ~8.1 & ~9.8 & 10.4 &        \\
  NGC 3507                 & SB(s)b     & 19.8 &      &      &        \\
  NGC 3607                 & Sa(s)0     & 19.9 & 21.2 &      &        \\
  NGC 3608                 & E2         & 23.4 & 21.3 &      &        \\
  NGC 3628                 & SAb        & ~7.7 &      &      &        \\
  NGC 3998                 & SA(r)0     & 14.0 & 13.1 &      &        \\
  NGC 4111                 & SA(r)0     & 17.0 & 13.9 &      &        \\
  NGC 4125                 & E6         & 24.2 & 22.2 &      &        \\
  NGC 4143                 & SAB(s)0    & 17.0 & 14.8 &      &        \\
  NGC 4261 (3C 270)        & E2-3       & 30.0 & 31.6 &      & 36.1   \\
  NGC 4278                 & E1-2       & ~9.7 & 14.9 &      &        \\
  NGC 4314                 & SB(rs)a    & ~9.7 &      &      &        \\
  NGC 4374 (M84, 3C 272.1) & E1         & 16.8 & 17.1 & 18.0 &        \\
  NGC 4438                 & SA(s)0/a   & 16.8 &      &      & 11.3   \\
  NGC 4457                 & SAB(s)0/a  & 17.4 &      &      & 10.7   \\
  NGC 4486 (M87, 3C 274)   & E0-1       & 16.8 & 14.9 & 15.8 &        \\
  NGC 4494                 & E1-2       & ~9.7 & 15.8 &      & 16.8   \\
  NGC 4548 (M91)           & SBb(rs)    & 16.8 & 17.9 &      & 15.0   \\
  NGC 4552 (M89)           & E          & 16.8 & 14.3 &      & 19.6   \\
  NGC 4579 (M58)           & SAB(rs)b   & 16.8 &      &      & 21.0   \\
  NGC 4594 (M104)          & SA(s)a     & ~9.2 & ~9.1 & ~9.6 & ~9.3   \\
  NGC 4636                 & E/S0       & 17.0 & 13.6 & 18.1 &        \\
  NGC 4736 (M94)           & (R)SA(r)ab & ~4.3 & ~4.8 & ~4.4 & ~4.7   \\
  NGC 5055 (M63)           & SA(rs)bc   & ~7.2 &      &      &        \\
  NGC 5866                 & S0         & 15.3 & 14.3 & 15.1 &        \\
  NGC 6500                 & SAab       & 39.7 &      &      &        \\
  NGC 7331                 & SA(s)b     & 14.3 & 12.2 &      & 14.5   \\
\enddata
\tablenotetext{a}{Host galaxy Hubble types were taken from
  the catalog of \citet{tully88}.}
\tablenotetext{b} {Distances from the catalog of \citet{tully88}.}
\tablenotetext{c} {Distances obtained using the surface brightness
fluctuation method. See \citet{tonry01}. Following \citet{jensen03},
we have corrected the distance modulus reported by \citet{tonry01} by
subtracting 0.16~mag.}
\tablenotetext{d} {Distances obtained using the planetary nebula
luminosity function method. See \citet{herrmann08} for NGC~4736
and compilation of \citet{ciardullo02} and references therein for 
other galaxies.}
\tablenotetext{e} {Distances obtained using a variety of different
methods, as follows: NGC~404, NGC~3031, and NGC~4736 using the tip of
the red giant branch (TRGB) method \citet{karachentsev02} \citep[the
distance to NGC~3031 obtained by the Cepheid method by][agrees exactly
with the TRGB distance]{freedman94}; NGC~4261, NGC~4552, and NGC 4636
using the fundamental plane relation \citep[see][]{gavazzi99};
NGC~4438, NGC~4457, and NGC~4579 using Tully-Fisher method
\citep[see][]{gavazzi99}; NGC~3031, NGC~4548 and NGC~7331 using the
Cepheid variable method \citep{freedman01,freedman94}.}
\end{deluxetable}

\subsection{Distances}\label{S:distances}

We have compiled distance measurements from the literature that are
based on a variety of techniques, which we include in
Table~\ref{T:hosts}. All 35 galaxies have distances cataloged in
\citet{tully88}, which were determined using a model for peculiar
velocities and assuming $H_0=75~{\rm km~s^{-1}~Mpc^{-1}}$. These are
listed in column (3) of Table~\ref{T:hosts}. For a significant
fraction of the galaxies in our collection, distance measurements
based on more direct techniques are available, which we also list in
Table~\ref{T:hosts}. More specifically, 24 galaxies have had their
distances determined via surface brightness fluctuations (SBF) by
\citet{tonry01}, which we list in column (4) of Table~\ref{T:hosts}
after making a systematic correction of $-0.16$~mag to the distance
modulus, following \citet{jensen03}.  In column (5) of
Table~\ref{T:hosts}, we list distances to 8 galaxies, determined via
the planetary nebula luminosity function (PNLF) method and drawn
mostly from the compilation of \citet{ciardullo02}, and references
therein. Distances determined by any other method are included in
column (6) of Table~\ref{T:hosts}. In this paper, we adopt the
distances from \citet{tully88} so that the luminosities we derive can
be compared directly to other quantities reported by \citet{ho97b},
who used the same distances (e.g. emission line luminosities). None of
the conclusions of this paper or \ptwo\ depend sensitively on the
distance. However, we emphasize that different applications (e.g.,
detailed study and modeling of the \seds) may require a more accurate
distance than that of \citet{tully88}. In such a case the data and
derived quantities we present here can be easily converted to a
different distance, via a simple scaling.

\subsection{Spectroscopic Classification and LINER Types}\label{S:types}

\begin{deluxetable}{lcccll}
\tabletypesize{\scriptsize}
\tablewidth{3.35in}
\tablecaption{Properties of LINER Nuclei\label{T:nuclei}}
\tablehead{
\colhead{} &
\colhead{Galactic} &
\colhead{} &
\colhead{} &
\multicolumn{2}{c}{LINER Type} \\
\colhead{} &
\colhead{$E(B-V)$\tablenotemark{a}} &
\colhead{$\sigma_{\star}$\tablenotemark{b}} &
\colhead{log} &
\multicolumn{2}{c}{\hrulefill} \\
\colhead{Galaxy} &
\colhead{(mag)} &
\colhead{(km~s$^{-1}$)} &
\colhead{$(\Mbh/{\rm M}_{\odot})$\tablenotemark{c}} &
\colhead{Ho\tablenotemark{d}} &
\colhead{Kewley\tablenotemark{e}} \\
\colhead{(1)} &
\colhead{(2)} &
\colhead{(3)} &
\colhead{(4)} &
\colhead{(5)} &
\colhead{(6)}  
}
\startdata
  NGC 0266 & 0.069 & ... & ... & L1       & L/L/L    \\
  NGC 0404 & 0.059 &  40 & 5.3 & L2       & L/C/L    \\
  NGC 1097 & 0.027 & 196 & 8.1 & L1       & L/L/S    \\
  NGC 1553 & 0.013 & 177 & 7.9 & L2/T2    & ?/L/?    \\
  NGC 2681 & 0.023 & 108 & 7.1 & L1       & L/L/H    \\
  NGC 3031 & 0.080 & 162 & 7.8 & S1.5/L1  & L/L/S    \\
  NGC 3169 & 0.031 & 163 & 7.8 & L2       & L/L/L    \\
  NGC 3226 & 0.023 & 193 & 8.1 & L1       & L/L/L    \\
  NGC 3379 & 0.024 & 207 & 8.2 & L2/T2    & L/L/L    \\
  NGC 3507 & 0.024 & ... & ... & L2       & L/L/H    \\
  NGC 3607 & 0.021 & 224 & 8.4 & L2       & L/L/S    \\
  NGC 3608 & 0.021 & 192 & 8.1 & L2/S2    & S/L/S    \\
  NGC 3628 & 0.027 & 171 & 7.9 & T2       & S/L/S    \\
  NGC 3998 & 0.016 & 305 & 8.4 & L1       & L/L/S    \\
  NGC 4111 & 0.015 & 148 & 7.6 & L2       & L/L/H    \\
  NGC 4125 & 0.019 & 227 & 8.4 & T2       & L/L/L    \\
  NGC 4143 & 0.013 & 214 & 8.3 & L1       & L/L/L    \\
  NGC 4261 & 0.018 & 309 & 8.7 & L2       & L/L/L    \\
  NGC 4278 & 0.029 & 261 & 8.6 & L1       & L/L/L    \\
  NGC 4314 & 0.025 & 117 & 7.2 & L2       & L/C/H    \\
  NGC 4374 & 0.040 & 308 & 9.2 & L2       & L/L/L    \\
  NGC 4438 & 0.028 & ... & ... & L1       & L/L/L    \\
  NGC 4457 & 0.022 &  96 & 6.9 & L2       & L/L/H    \\
  NGC 4486 & 0.022 & 333 & 9.5 & L2       & L/L/L    \\
  NGC 4494 & 0.021 & 145 & 7.6 & L2       & L/L/L    \\
  NGC 4548 & 0.038 & 144 & 7.6 & L2       & L/L/L    \\
  NGC 4552 & 0.041 & 203 & 8.2 & T2       & L/L/L    \\
  NGC 4579 & 0.041 & 165 & 7.8 & L1       & L/L/L    \\
  NGC 4594 & 0.051 & 241 & 9.0 & L2       & L/L/L    \\
  NGC 4636 & 0.028 & 203 & 8.2 & L1       & L/L/L    \\
  NGC 4736 & 0.018 & 112 & 7.1 & L2       & L/L/L    \\
  NGC 5055 & 0.018 & 108 & 7.1 & T2       & L/L/H    \\
  NGC 5866 & 0.013 & 159 & 7.7 & T2       & S/L/S    \\
  NGC 6500 & 0.090 & 214 & 8.3 & L2       & L/C/L    \\
  NGC 7331 & 0.091 & 138 & 7.5 & T2       & S/L/S    \\
\enddata
\tablenotetext{a}{Reddening caused by the ISM of the Milky Way; from
\citet{schlegel98}.}
\tablenotetext{b}{The stellar velocity dispersion, taken from the
  Hypercat database, with the following exceptions: NGC~404, NGC~4278,
  NGC~4314, NGC~4374, NGC~4579, NGC~4736, NGC~5055, NGC~6500 from
  \citet{barth02}; NGC~1097 from \citet{lewis06}.}
\tablenotetext{c}{The log of the black hole mass in
  M$_{\odot}$. Derived from stellar velocity dispersions using
  equation (\ref{Q:bhmass}), with the following exceptions: for
  NGC~3031 it was derived from stellar and gas kinematics
  \citep{bower00,devereux03}, while for NGC~4261 and NGC~4374 it was
  derived from gas kinematics \citep{ferrarese96,bower98}.}
\tablenotetext{d}{Spectroscopic classification according to
  \citet{ho97b}, with the exception of NGC~1097 and NGC~1553. The
  LINER types for these two galaxies were taken from
  \citet{phillips84} and \citet{phillips86}, respectively.  L1=LINER
  with a broad H$\alpha$ line, L2=LINER without a broad H$\alpha$
  line, T2=intermediate emission line ratios between LINER and
  \ion{H}{2} region, S=Seyfert; combinations indicate intermediate
  line ratios between two classes.}
\tablenotetext{e} {Spectroscopic classification based on the criteria
  of \citet{kewley01}, \citet{kauffmann03}, and \citet{kewley06}. The
  three designations refer to the location of the object in the
  [\ion{O}{1}]/H\a\ $vs$ [\ion{O}{3}]/H\b, [\ion{N}{2}]/H\a\ $vs$
  [\ion{O}{3}]/H\b, and [\ion{S}{2}]/H\a\ $vs$ [\ion{O}{3}]/H\b\
  diagrams, respectively. L=LINER, S=Seyfert, H=\ion{H}{2} region, C=
  ``composite,'' i.e., intermediate between LINERs or Seyferts and
  \ion{H}{2} regions in the [\ion{N}{2}]/H\a\ $vs$ [\ion{O}{3}]/H\b
  diagram.}
\end{deluxetable}

In column (5) of Table~\ref{T:nuclei} we list the LINER types
according to \citet{ho97b}, which are based on the relative
intensities of the narrow emission lines. The objects included in
our sample are classified in this scheme either as pure LINERs
(denoted by L), or as transition objects, with diagnostic line ratios
intermediate between LINERs and \ion{H}{2} regions. In about 1/3 of
the objects in this sample, there are detectable broad wings on the
H\a\ line \citet{ho97c}; these are termed ``type 1.9'' LINERs and are
identified with a ``1'' in column (5) of Table~\ref{T:nuclei}, while
all other objects are labeled with a ``2.''  If there was any
ambiguity or uncertainty in the classification, both possible classes
are listed. The nucleus of M81 has an uncertain classification; it can
be either a LINER or a Seyfert. Since the work of \cite{ho97b}, more
recent classification schemes using the same diagnostic line ratios
have become available, such as those of \citet{kewley01},
\citet{kauffmann03}, and \citet{kewley06}. Thus, we have applied these
classification schemes to the relative intensities of diagnostic lines
measured by \citet{ho97b} and report the outcome of this exercise in
column~(6) of Table~\ref{T:nuclei}. We list the location of each
nucleus in the [\ion{O}{1}]/H\a\ $vs$ [\ion{O}{3}]/H\b,
[\ion{N}{2}]/H\a\ $vs$ [\ion{O}{3}]/H\b, and [\ion{S}{2}]/H\a\ $vs$
[\ion{O}{3}]/H\b\ diagrams, with L denoting the LINER region, S
denoting the Seyfert region, H denoting the \ion{H}{2} region, and C
denoting the ``composite'' region in the [\ion{N}{2}]/H\a\ $vs$
[\ion{O}{3}]/H\b\ diagram (intermediate between LINERs or Seyferts and
\ion{H}{2} regions). Using these criteria, six objects are classified
as ``S/L/S,'' however they fall very close to the Seyfert-LINER
boundary in the [\ion{O}{1}]/H\a\ $vs$ [\ion{O}{3}]/H\b\ and
[\ion{S}{2}]/H\a\ $vs$ [\ion{O}{3}]/H\b\ diagrams. Similarly, NGC~4314
is classified as ``L/C/H'' but it falls very close to the
Composie-LINER and \ion{H}{2}-LINER boundaries in [\ion{N}{2}]/H\a\
$vs$ [\ion{O}{3}]/H\b, and [\ion{S}{2}]/H\a\ $vs$ [\ion{O}{3}]/H\b\
diagrams, respectively.

The mix of LINER types in our sample is dictated by the availability
of data. Therefore, there may some biases inherited from the surveys
from which we adopted the data. More specifically, the surveys we have
relied on targeted radio-bright, UV-bright, and X-ray bright
objects. Thus, we find that ``transition'' objects \citep[according
to][]{ho97b} are under-represented in our sample relative to their
number in the survey of \citet{ho97a}, while ``type 1.9'' LINERs are
over-represented.  We also note that our search for data, although
extensive, was not exhaustive, with the result that we may have
overlooked a few objects. It is not clear whether the relative number
of LINER types in our sample should have an effect on our
conclusions. If the properties of the AGN are related to the LINER
class, then the composition of the sample may influence any average
properties derived from it. We attempt to assess such effects at the
end of \S\ref{S:aox} using our estimated bolometric luminosities.

\subsection{Stellar Velocity Dispersions and Black Hole Masses}\label{S:masses}

The stellar velocity dispersions of the sample galaxies were taken
from the Hypercat database\footnote{\tt
http://www-obs.univ-lyon1.fr/hypercat}, with the following exceptions:
the values for NGC~404, NGC~4278, NGC~4314, NGC~4579,
NGC~4736, NGC~5055, NGC~6500 are from \citet{barth02}, while the value
for NGC~1097 is from \citet{lewis06}.  The black hole masses were
estimated from the stellar velocity dispersions via the
$\Mbh$--$\sigma_{\star}$ relationship
\citep{ferrarese00,gebhardt00,tremaine02}, namely,
\begin{equation}
\log\left({M/\Msol}\right) = \alpha + \beta\;
\log\left({\sigma_{\star}/\sigma_0}\right)\; , 
\label{Q:bhmass}
\end{equation}
where $\alpha = 8.13 \pm 0.06$, $\beta = 4.02 \pm 0.32$, and $\sigma_0
= 200~\kms$ \citep{tremaine02}. For the following galaxies we adopt
the black hole masses measured from spatially resolved stellar and/or
gas kinematics: NGC~3031 \citep{bower00,devereux03}, NGC 3998
\citep{defrancesco06} NGC~4261 \citep{ferrarese96}, NGC~4374
\citep{bower98}, NGC~4486 \citep{machetto97}, and NGC~4594
\citep{kormendy96}.
In the case of NGC~3031 separate measurements from stellar kinematics,
gas kinematics, and the central stellar velocity dispersion give
nearly identical values (within 5\%). In the case of NGC~4261, the
black hole mass from gas kinematics differs from that obtained from
the central stellar velocity dispersion by a factor of only 1.6.  In
seven cases where the stellar velocity dispersion is reported both in
the Hypercat database and by \citet{barth02}, the black hole masses
are within a factor 1.6 or less from each other.  There are four large
discrepancies, as follows: in NGC~404 where the stellar velocity
dispersions reported in the Hypercat database and by \citet{barth02}
lead to black hole masses that differ by a factor of 3.6, while in
NGC~3998, NGC~4486, and NGC~4594 mass determinations from the central
stellar velocity dispersion and spatially resolved stellar or gas
kinematics differ by factors between 2.7 and 3.5. Considering all cases
with multiple determinations of the black hole mass, the values of
$\Delta\log(M/{\rm M}_\odot)$ are evenly distributed about zero with a
standard deviation of 0.34 (amounting to a factor of approximately
2).
There are three galaxies for which we could not estimate the black
hole masses because we could not find the necessary data: NGC~266,
NGC~3507, and NGC~4438.  The distribution of black hole masses is
shown in Figure~\ref{F:hosts}d. Their values span the range $5.3 <
\log\left({M/\Msol}\right) < 9.5$.

\section{Spectral Energy Distributions}\label{S:sed}

\subsection{Data Compilation}\label{S:data}

\begin{figure*}
\centerline{
            \includegraphics[scale=0.55,angle=0]{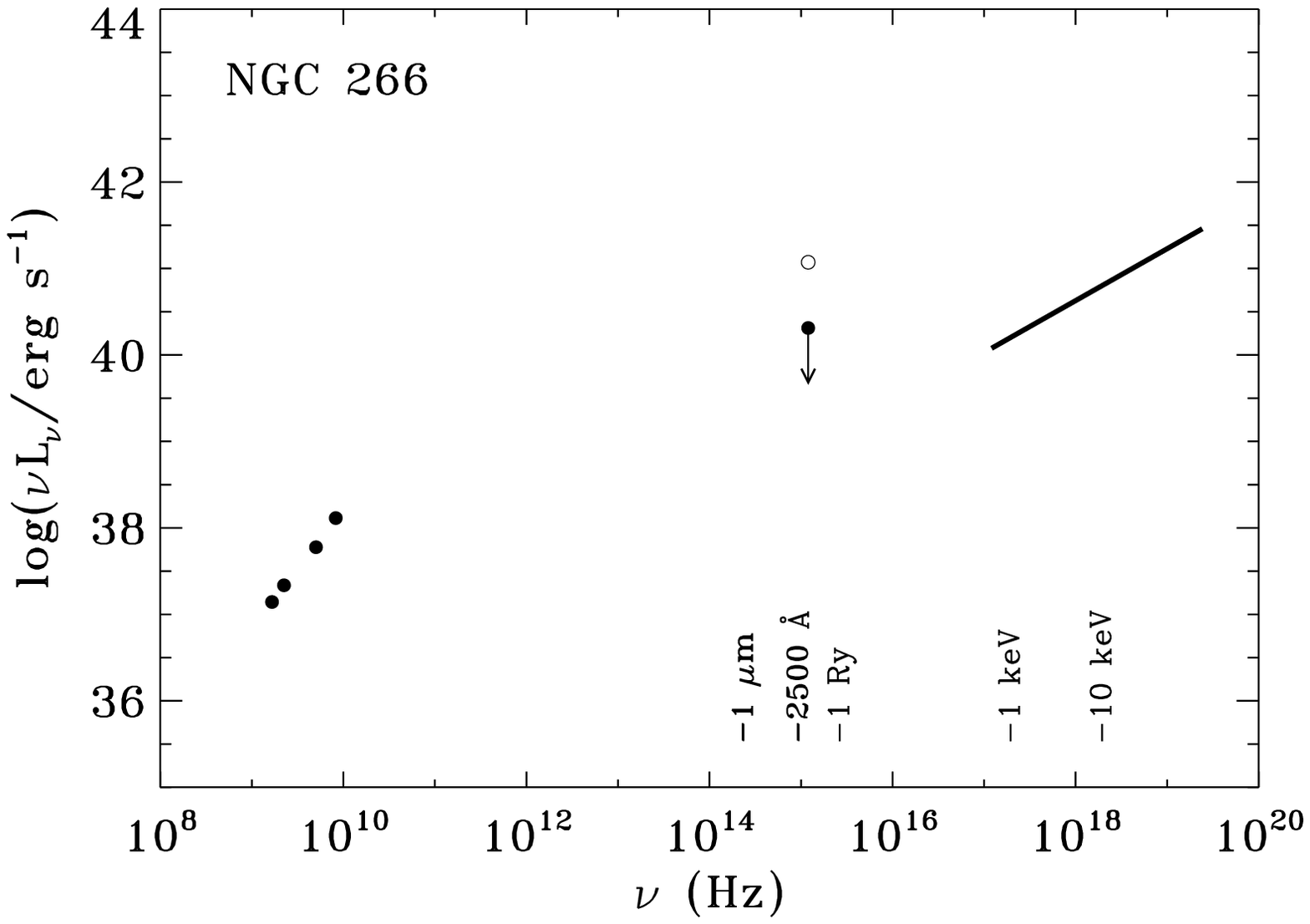} 
\hskip -0.3truein                                        
            \includegraphics[scale=0.55,angle=0]{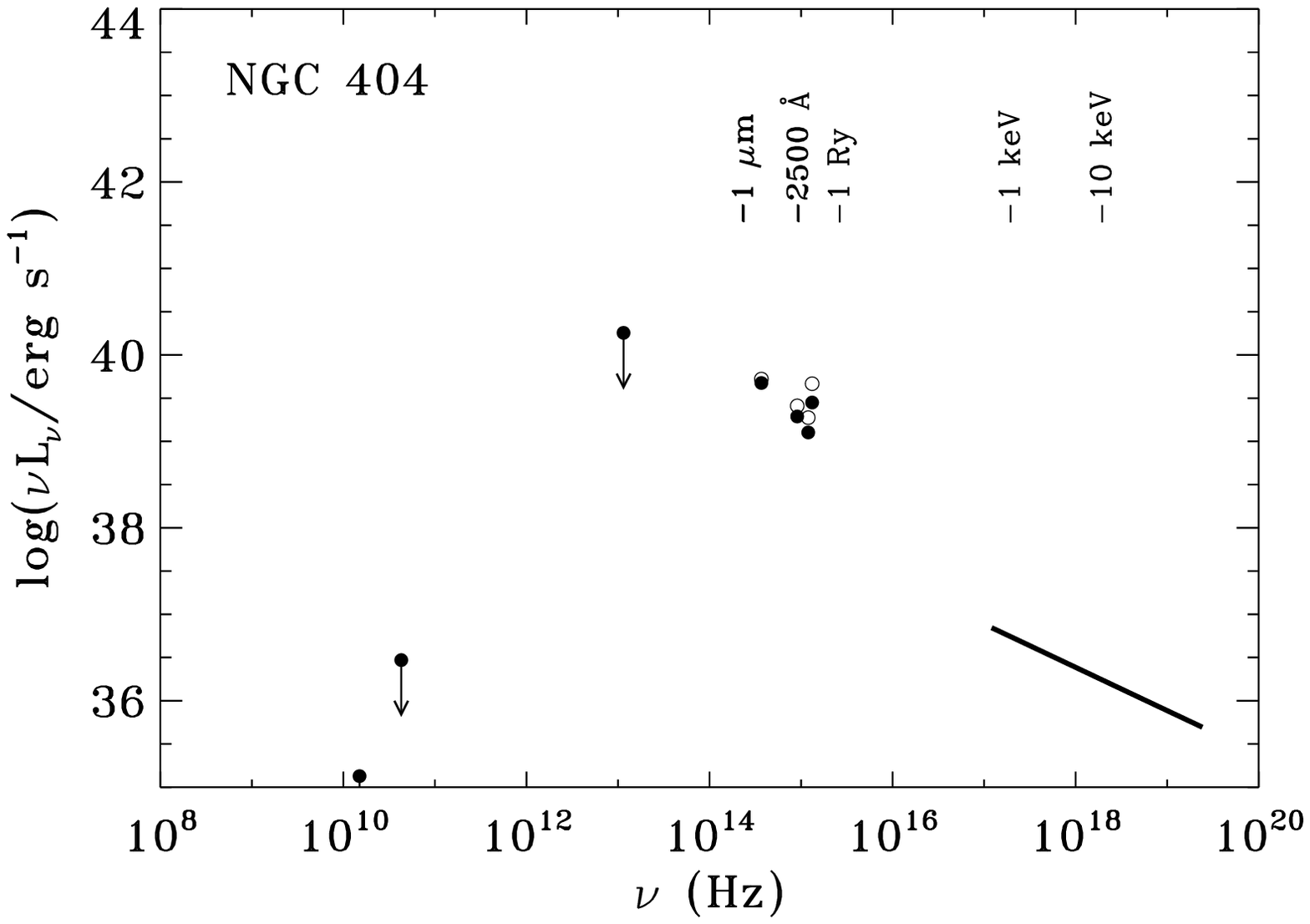} 
}                                                        
\centerline{                                             
            \includegraphics[scale=0.55,angle=0]{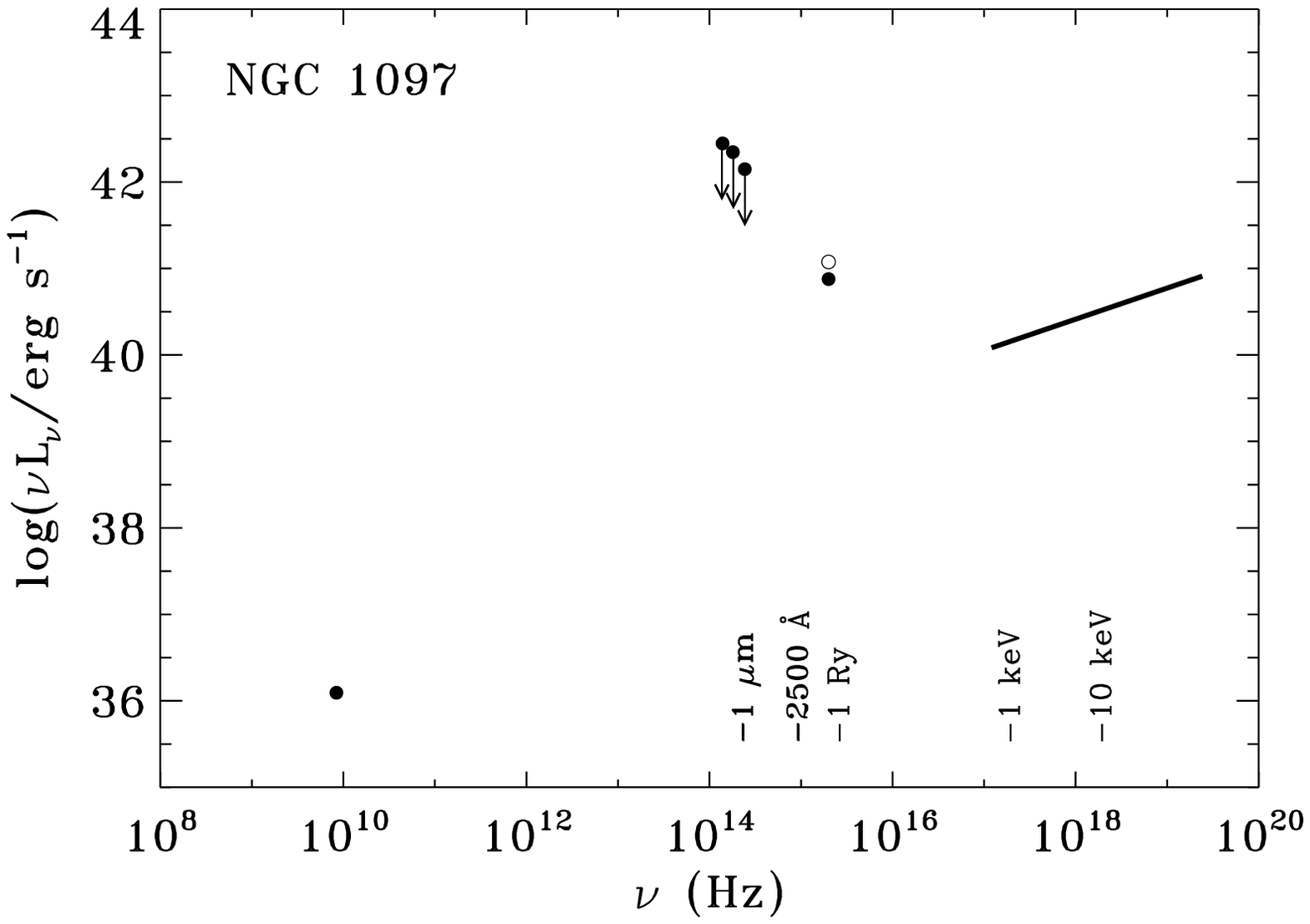} 
\hskip -0.3truein                                        
            \includegraphics[scale=0.55,angle=0]{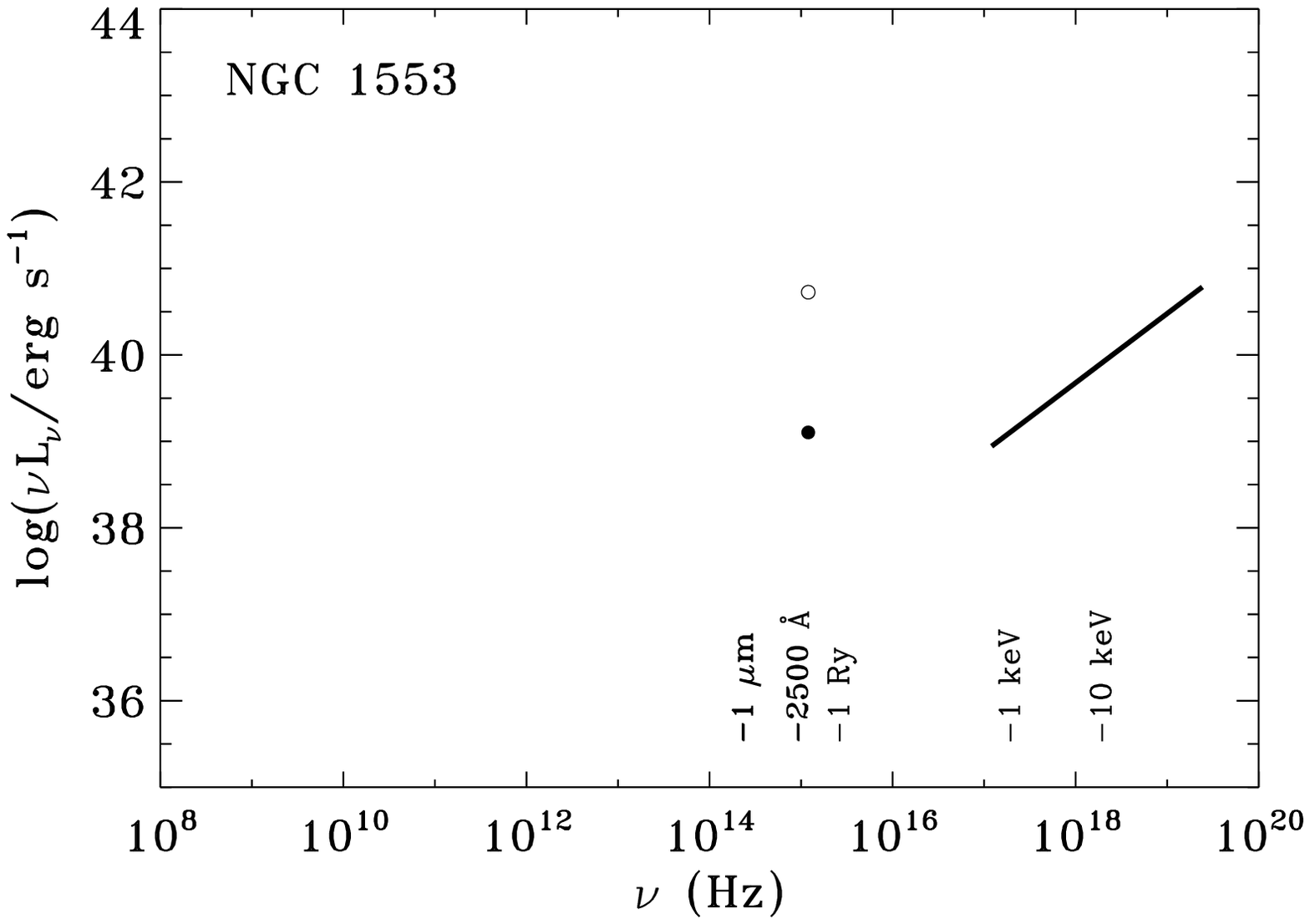} 
}
\centerline{
\hskip 0.55 truein
\vbox to 2.5 truein {\hsize 3 truein
\caption{The spectral energy distributions of individual galaxies in
  our sample. The data points represent measurements in a moderately
  narrow band ($\Delta\nu/\nu\ls 0.2$), centered at a specific
  frequency. Whenever multiple measurements were available at
  approximately the same frequency, these were averaged together and
  their standard deviation is represented as a vertical error bar (see
  discussion in \S\ref{S:data} of the text).  The filled circles
  represent measurements without extinction corrections, while the
  open circles are the same measurements after extinction corrections
  (see details in \S\ref{S:extinction} of the text).  The thick solid
  lines represent the power-law component of the best-fitting model to
  the 0.5--10 keV X-ray spectrum (already corrected for extinction and
  extrapolated to 100~keV, for reference). In the case of NGC~4261, we
  do not plot the optical points after extinction corrections; see the
  discussion in \S\ref{S:objects} of the text. \\ \\
  THE COMPLETE FIGURE WILL BE AVAILABLE IN THE ELECTRONIC VERSION OF 
  THE JOURNAL.
  \label{F:indsed}
  \vfill
}
}
            \includegraphics[scale=0.55,angle=0]{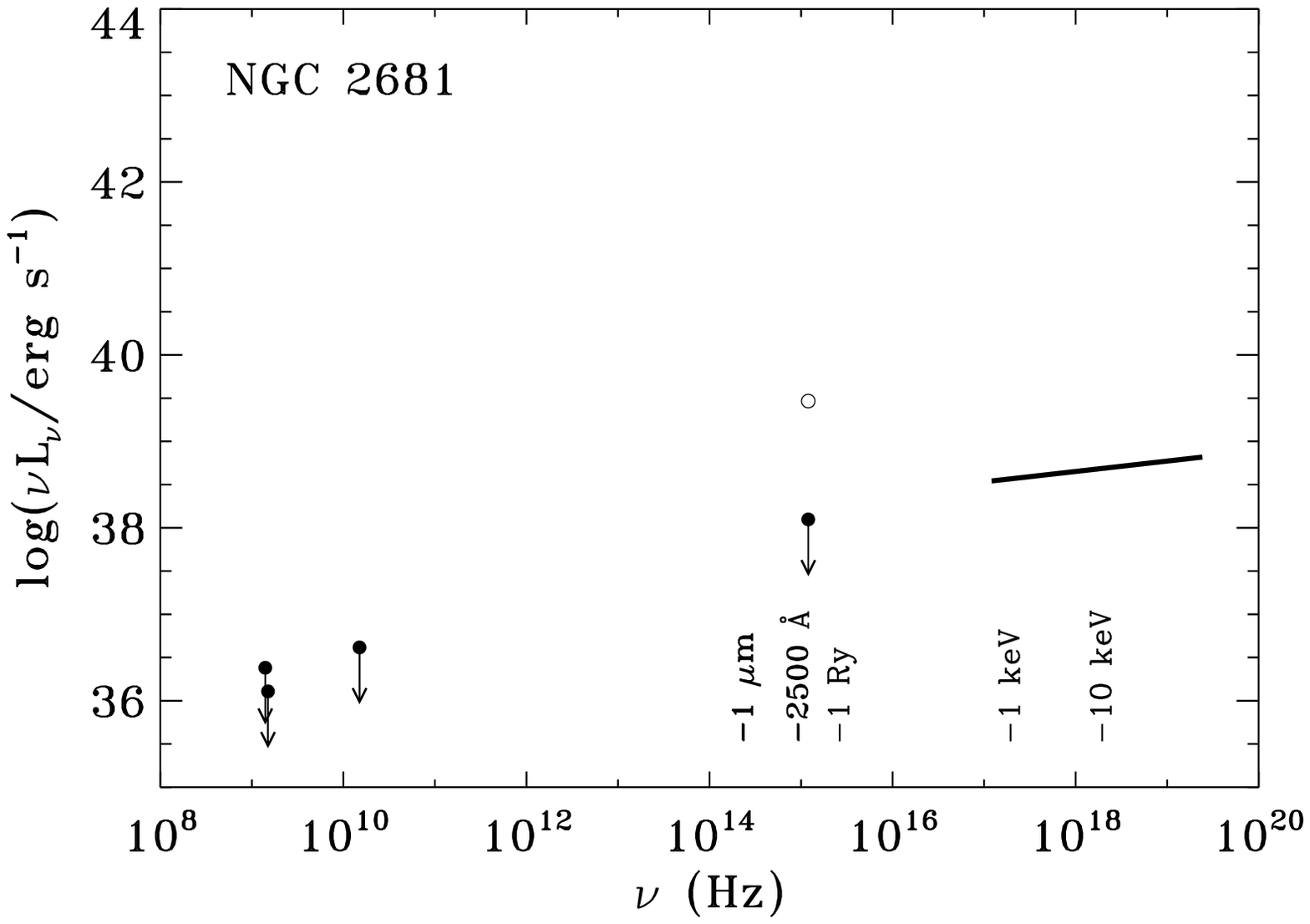} 
}
\end{figure*}

In Table~\ref{T:indsed} we present the data making up the \seds\ of
individual galaxies in the form of monochromatic luminosity versus
frequency (i.e., $\nu L_{\nu}~vs~\nu$). These data were taken from
sources in the literature, as listed in Table~\ref{T:indsed}. We
include in this table monochromatic luminosities with and without
corrections for extinction (see discussion of extinction corrections
below). As we noted in \S\ref{S:sample}, we include primarily
measurements made at high spatial resolution ($<1$\asec); in some
cases observations at lower spatial resolution are used, but the
resulting measurements are treated as upper limits. Upper limits from
non-detections are also listed in this table. In cases where UV
(2500~\AA) measurements or limits are not available but X-ray
measurements or upper limits are, we inferred UV upper limits from the
2~keV monochromatic luminosities and the assumption that the
optical-to-X-ray spectral index is $\aox < 1.5$. We justify this
assumption in \S\ref{S:aox}, below.  Column (3), labeled ``Observed
\nLn,'' gives the monochromatic luminosity obtained from the observed
flux density without any corrections. Columns (4) and (5), labeled
``Corrected \nLn,'' give the monochromatic luminosity after minimum
\citep[following][]{calzetti94} and maximum
\citep[following][]{seaton79} extinction corrections. We discuss these
corrections in detail in \S\ref{S:extinction} below. The individual
\seds\ are shown graphically in Figure~\ref{F:indsed}, in a separate
panel for each galaxy; the ``Observed \nLn'' is represented by filled
points, with arrows denoting upper limits, while open points show
\nLn\ after the maximum extinction correction.

\begin{deluxetable*}{llrrrcll}
\tabletypesize{\scriptsize}
\tablewidth{7in}
\tablecaption{Spectral Energy Distributions of Individual LINERs\label{T:indsed}}
\tablehead{
\colhead{} &
\colhead{} &
\colhead{Observed} &
\multicolumn{2}{c}{Corrected $\nu\; L_{\nu}$ (erg s$^{-1}$)\tablenotemark{a}} &
\colhead{} &
\colhead{} &
\colhead{} \\
\colhead{} &
\colhead{$\nu$} &
\colhead{$\nu\; L_{\nu}$} &
\multicolumn{2}{c}{\hrulefill} &
\colhead{} &
\multicolumn{2}{c}{References\tablenotemark{c}} \\
\colhead{\hbox to 0.5truein{\hfil Galaxy\hfil}} &
\colhead{\hbox to 0.75truein{\hfil (Hz)\hfil}} &
\colhead{\hbox to 0.75truein{\hfil (erg s$^{-1}$)\hfil}} &
\colhead{\hbox to 0.75truein{\hfil Calzetti (min)\hfil}} &
\colhead{\hbox to 0.75truein{\hfil Seaton (max)\hfil}} &
\colhead{\hbox to 0.75truein{\hfil $\sigma/\langle\nu\; L_{\nu}\rangle$\tablenotemark{b}\hfil}} &
\multicolumn{2}{c}{and Notes} \\
\colhead{(1)} &
\colhead{(2)} &
\colhead{(3)} &
\colhead{(4)} &
\colhead{(5)} &
\colhead{(6)} &
\multicolumn{2}{c}{(7)}  
}
\startdata                                                                                                        
  NGC  266 & $1.700\times 10^{ 9}$ & $ 1.43\times 10^{37}$ & \dots                 & \dots                 &        &  1         & \tablenotemark{d} \\   
           & $2.300\times 10^{ 9}$ & $ 2.14\times 10^{37}$ & \dots                 & \dots                 &        &  1         & \tablenotemark{d} \\   
           & $5.000\times 10^{ 9}$ & $ 6.06\times 10^{37}$ & \dots                 & \dots                 &        &  1         & \tablenotemark{d} \\   
           & $8.400\times 10^{ 9}$ & $ 1.41\times 10^{38}$ & \dots                 & \dots                 &        &  1         & \tablenotemark{d} \\   
           & $1.199\times 10^{15}$ & $<2.05\times 10^{40}$ & $<1.18\times 10^{41}$ & $<2.88\times 10^{41}$ &        &            & \tablenotemark{e} \\   
           & $1.210\times 10^{17}$ & $ 1.20\times 10^{40}$ & \dots                 & \dots                 &        &  2         & \tablenotemark{~} \\   
           & $2.420\times 10^{17}$ & $ 1.81\times 10^{40}$ & \dots                 & \dots                 &        &  2         & \tablenotemark{~} \\   
           & $2.420\times 10^{18}$ & $ 7.22\times 10^{40}$ & \dots                 & \dots                 &        &  2         & \tablenotemark{~} \\   
\hline
  NGC  404 & $1.500\times 10^{10}$ & $<1.34\times 10^{35}$ & \dots                 & \dots                 &        &  3         & \tablenotemark{f} \\   
           & $4.286\times 10^{10}$ & $<2.95\times 10^{36}$ & \dots                 & \dots                 &        &  4         & \tablenotemark{f} \\   
           & $1.149\times 10^{13}$ & $<1.80\times 10^{40}$ & \dots                 & \dots                 &        &  5         & \tablenotemark{g} \\   
           & $3.686\times 10^{14}$ & $ 4.74\times 10^{39}$ & $ 4.74\times 10^{39}$ & $ 5.26\times 10^{39}$ &        &  6         & \tablenotemark{~} \\   
           & $9.085\times 10^{14}$ & $ 1.94\times 10^{39}$ & $ 2.06\times 10^{39}$ & $ 2.58\times 10^{39}$ &        &  7         & \tablenotemark{h} \\   
           & $1.199\times 10^{15}$ & $ 1.27\times 10^{39}$ & $ 1.39\times 10^{39}$ & $ 1.88\times 10^{39}$ &        &  7         & \tablenotemark{h} \\   
           & $1.322\times 10^{15}$ & $ 2.82\times 10^{39}$ & $ 3.12\times 10^{39}$ & $ 4.65\times 10^{39}$ &        &  8         & \tablenotemark{~} \\   
           & $1.210\times 10^{17}$ & $ 7.00\times 10^{36}$ & \dots                 & \dots                 &        &  9         & \tablenotemark{~} \\   
           & $2.420\times 10^{17}$ & $ 4.95\times 10^{36}$ & \dots                 & \dots                 &        &  9         & \tablenotemark{~} \\   
           & $2.420\times 10^{18}$ & $ 1.57\times 10^{36}$ & \dots                 & \dots                 &        &  9         & \tablenotemark{~} \\   
\hline
\noalign{
\medskip
\hfil THE COMPLETE TABLE WILL BE AVAILABLE IN THE ELECTRONIC VERSION OF THE JOURNAL \hfil}
\enddata
\tablenotetext{a}{The monochromatic luminosities from 0.1 to 1$\mu$m
                  after correction for extinction.  The minimum
                  correction employs the starburst extinction law of
                  \citet{calzetti94}, while the maximum correction
                  corresponds to the Milky Way law of
                  \citet{seaton79}.  Details are given in \S{S:data}
                  of the text. The ``observed'' X-ray luminosities
                  already have this correction built in. The distances
                  used are those of \citet{tully88}; see
                  Table~\ref{T:hosts} and the discussion in
                  \S\ref{S:distances} of the text.}
\tablenotetext{b}{The fractional dispersion in the monochromatic
                  luminosity in cases where a number of measurements
                  were averaged together. See \S\ref{S:data} of the
                  text for details.}
\tablenotetext{c}{{\it References. --}
 (1)  \citet{doi05a};
 (2)  \citet{terashima03};
 (3)  \citet{nagar05};
 (4)  \citet{nagar00};
 (5)  \citet{satyapal04};
 (6)  \citet{chiaberge05};
 (7)  \citet{maoz05};
 (8)  \citet{maoz95};
 (9)  \citet{eracleous02}.
}
\tablenotetext{d}{See detailed notes on this object in
  \S\ref{S:objects} of the text.}
\tablenotetext{e}{Upper limit to $\nu L_{\nu}(2500\;{\rm\AA})$ derived
  by assuming $\alpha_{\rm ox}=1.5$. See \S\ref{S:aox} of the text for
  details.}
\tablenotetext{f}{This limit is a result of a non-detection.}
\tablenotetext{g}{This limit is a result of contamination of the
  source by other, neighboring sources; the observations were taken
  through a large aperture.}
\tablenotetext{h}{The observed UV flux most likely originates in hot
  stars in the immediate vicinity of the nucleus, not in the AGN. See
  the discussion in \S\ref{S:objects} of the text. The values of
  \aox\ implied by the UV flux is extremely high for such a
  low-luminosity AGN thus, the UV flux listed here can be taken as a
  {\it generous} upper limit to the UV flux of the AGN.}
\end{deluxetable*}

With the exception of the X-ray data, measurements were made in
relatively narrow bands ($\Delta\nu/\nu\ls 0.2$), centered at a
specific frequency. Whenever multiple measurements were available at
approximately the same frequency, these were averaged together and
their fractional standard deviations, $\sigma/(\nu L_{\nu})$, are
given in a separate column of Table~\ref{T:indsed}. These variations
can be caused by intrinsic variability of the source
\citep[specifically in the near UV; cf,][]{maoz05} or differences in
spatial resolution. In the former case the temporal fluctuations are
often less than 10\% on time scales of about a year. However,
\citet{maoz05} have also found some examples of larger variations (up
to 50\%) on these short time scales. More importantly, they have 
found that on time scales of several years to a decade the amplitude
of the fluctuations can reach a factor of a few.  In the case of
near-IR and radio data, the fluctuations can be as large as
$\sim60$\%. We have also encountered an example of even more extreme
variability in the radio band, which we discuss briefly in
\S\ref{S:objects}.

The X-ray data are in the form of broad-band spectra, typically
spanning the energy range 0.5--10~keV. Many of the X-ray spectra can
be fitted with simple power-law models, modified by interstellar
photoelectric absorption in nearly neutral gas. In some cases, more
complex models are needed, consisting of a power law plus optically
thin emission from a thermal plasma \citep[see, for example,][and
references therein]{eracleous02,flohic06}. In such cases, the
power-law component is attributed to the AGN and the thermal plasma
component is ascribed to spatially-unresolved, circumnuclear
emission. Thus, we have taken the power-law component from such models
to represent the emission from the AGN. In all cases, we adopt the
power-law model as a convenient parameterization of the data. This
model describes the intrinsic photon energy spectrum (number of emitted
photons per unit energy interval) as $N(E)= N_0 (E/E_0)^{-\Gamma}$,
where $E_0$ is a fiducial energy (typically $E_0=1$~kev), $\Gamma$ is
known as the ``photon index,'' and $N_0=N(E_0)$ is the ``normalization
constant'' (with units of ${\rm cm^{-2}~s^{-1}~keV^{-1}}$).  With this
convention, and keeping in mind that $f_{\nu}(E_0) = h\; N_0\; E_0$
(where $h$ is Planck's constant), we can write the flux density
spectrum as
$$
f_{\nu} (E) = f_{\nu} (E_0) \left({E\over E_0}\right)^{1-\Gamma}
\hbox to 1.5 truein{\hss}
$$
\begin{equation}
= 0.663 \left({N_0\over 1\;{\rm cm^{-2}\; s^{-1}\;keV^{-1}}}\right)
\left({E\over 1\;{\rm keV}}\right)^{1-\Gamma}~{\rm mJy}\; .
\label{Q:plaw}
\end{equation}
The parameters describing the power-law X-ray spectrum ($\Gamma$ and
$N_0$, for $E_0=1$~kev), as well as the equivalent hydrogen column
density ($N_{\rm H}$; inferred from fitting the model to the data) are
listed in Table~\ref{T:XUV}. Using these parameters we have calculated
the values of \nLn\ at 0.5, 1, and 10~keV, which we list in
Table~\ref{T:indsed} (the three highest-frequency values for each
object). We note that the AGNs in LINERs are typically detected at 0.5
and 1~keV, and often also at 10~keV, so it is fair to use the models
to evaluate \nLn\ at these energies. In Figure~\ref{F:indsed} we plot
the X-ray component of the \sed\ as a thick solid line between 0.5 and
10~keV (already corrected for extinction) and we extrapolate it to
100~keV, although only for reference.

In a number of cases, the X-ray emission from the AGN is not detected,
but an upper limit is available from the observations. These limits
are expressed in Table~\ref{T:XUV} as upper limits on the
normalization of the X-ray spectrum, for an assumed value of a photon
index of $\Gamma=1.8$.\footnote{This value of $\Gamma$ is the median
value found in the large sample of \citet{flohic06}. The derived
limits are not very sensitive to the value of $\Gamma$ \citep[see the
discussion in][and footnote (a) in Table~\ref{T:XUV}]{flohic06}. For
example, changing $\Gamma$ by $\pm 0.2$ changes the flux by only
$\sim10$\%.} In the case of one galaxy, NGC~3379, the normalization of
the X-ray spectrum is not an upper limit but it was derived under the
assumption of $\Gamma=1.8$, because of the low S/N of the X-ray
spectrum. The monochromatic luminosities of these objects given in
Table~\ref{T:indsed} are also identified as upper limits.

\subsection{Extinction Corrections}\label{S:extinction}

\begin{figure}
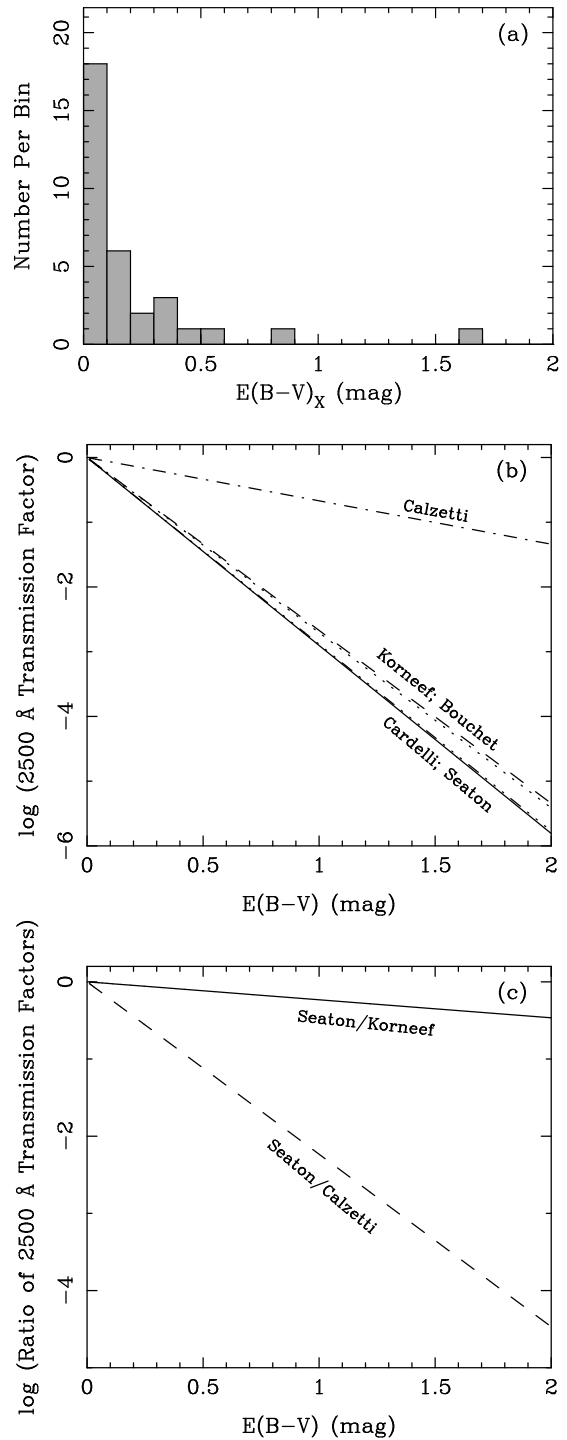

\centerline{\includegraphics[scale=0.42,angle=0]{f3a.eps}} 
\bigskip
\centerline{\includegraphics[scale=0.445,angle=0]{f3b.eps}} 
\centerline{\includegraphics[scale=0.445,angle=0]{f3c.eps}} 
\bigskip
\caption{Reddening values and extinction corrections for 33 of the 35
  objects in our sample (NGC~3169 and NGC~4261 are excluded because
  the column density is so large that they should be invisible in the
  UV). (a) The distribution of the values of $E(B-V)_{\rm X}$, the
  reddening derived from the X-ray column density (listed in
  Table~\ref{T:XUV}). As this histogram shows, 31/33 objects have
  $E(B-V)_{\rm X}<0.6$ and 22/33 objects have $E(B-V)_{\rm
    X}<0.2$. (b) The fraction of transmitted flux at 2500~\AA\ for the
  five extinction laws that we have explored, with line styles as
  follows (top to bottom): dot-dashed for the Starburst law of
  \citet{calzetti94}, dashed for the LMC law of \citet{korneef81},
  dotted for the LMC law of \citet{bouchet85}, and triple-dot-dashed
  and solid for the Milky Way laws of \citet{cardelli89} and
  \citet{seaton79}, respectively.  (c) Comparison of extinction laws
  via the ratios of 2500~\AA\ transmission factors. The
  ``Seaton/Calzetti'' ratio represents the full range of possible
  extinction corrections, while the ``Seaton/Korneef'' ratio
  represents the range of of possible non-starburst corrections.
\label{F:extinction}}
\end{figure}

To facilitate extinction corrections, we have listed the values of the
color excess (or reddening), $E(B-V)$, associated with the Galactic
ISM \citep*[taken from][]{schlegel98} in Table~\ref{T:nuclei}. The
model fits to the X-ray spectra yield a value for the {\it total}
equivalent hydrogen column density ($N_{\rm H}$, listed in
Table~\ref{T:XUV}). These models typically assume the photoelectric
absorption cross-sections of \citet{morrison83}, who adopted the
elemental abundances of \citet{anders82}. In Table~\ref{T:XUV}, we
also list the corresponding value of the {\it total} color excess,
derived from the following relation between the visual extinction,
$A_{\rm V}$, and the hydrogen column density \citep[from][applicable
to the Milky Way]{predhl95}
\begin{equation}
N_{\rm H}/A_{\rm V} = 1.79\times 10^{21}~\rm{cm^{-21}~mag^{-1}}\; ,
\label{Q:extinction}
\end{equation}
and assuming that $R_{\rm V}\equiv A_{\rm V}/E(B-V)=3.2$. As expected,
in most cases the {\it total} reddening, $E(B-V)_{\rm X}$
corresponding to the hydrogen column determined from the X-ray
spectra, is larger than the reddening caused by the Galactic ISM.  In
two cases, NGC~404 and NGC~3031, $E(B-V)$ exceeds $E(B-V)_{\rm X}$ by
0.019 and 0.008 magnitudes respectively, which is well within the
uncertainties of the column densities derived from X-ray observations.
In Figure~\ref{F:extinction}a, we show the distribution of values of
$E(B-V)_{\rm X}$ of the objects in our sample to illustrate that
$E(B-V)_{\rm X} < 0.6$ in about 31/33 of the objects and $E(B-V)_{\rm
X} < 0.2$ in about 22/33 of the objects (we exclude 2 objects with
extremely high extinction, in which no meaningful correction can be
made).  The values of $E(B-V)_{\rm X}$ should be regarded with caution
because of the assumptions involved in deriving them. Most uncertain
is the assumption of a Galactic gas-to-dust ratio, which is inherent
in the relation between the hydrogen column density and the visual
extinction given in equation~(\ref{Q:extinction}). Furthermore, there
is also the possibility that the lines of sight to the UV and X-ray
sources do not pass through exactly the same column of absorbing
material.

We applied extinction corrections to the observed monochromatic
luminosities as follows: Using the values of the {\it total}
reddening, $E(B-V)_{\rm X}$, for each galaxy we computed the
extinction corrections for points in the \sed\ between 0.1 and
1~$\mu$m using the following five extinction laws: the Milky Way
extinction laws of \citet{seaton79} and \citet*{cardelli89}, the Large
and Small Magellanic Cloud extinction laws of \citet{korneef81} and
\citet{bouchet85}, respectively, and the starburst galaxy extinction
law of \citet*{calzetti94}. 

All of the Milky Way and Magellanic Cloud laws agree well with each
other in the near-IR, optical, and near-UV bands but not in the far UV
\citep[see, for example, Figures~1 and 21 in][]{calzetti94}. The
differences are most pronounced in the far-UV band, at $\lambda <
1500$~\AA.  In the near-UV the Milky Way and LMC extinction curves are
not monotonic (because of the ``2200~\AA\ bump'') with the result that
the curves cross each other. Thus, the law that gives the highest
extinction correction changes with wavelength and also with the value
of $E(B-V)_{\rm X}$. To illustrate the effects of and differences
between the above extinction laws, we plot in
Figures~\ref{F:extinction}b and \ref{F:extinction}c, respectively, the
transmission factor (i.e., the fraction of flux that is transmitted)
at 2500~\AA\ according to each of the extinction laws and the
relations between the different transmission factors. The starburst
extinction law of \citet{calzetti94} differs considerably from the
other four laws.  Considering the Milky Way and Magellanic Cloud
extinction laws, we find that for $E(B-V)_{\rm X} < 0.2$ (22/33 of our
objects) the transmission factors range between 0.29 and 0.32 and
for $E(B-V)_{\rm X} < 0.5$ (31/33 of our objects) the the transmission
factors range between 0.02 and 0.03.

In the end we adopted the \citet{seaton79} law, which leads to the
largest corrections. We note, however, that if we had adopted any
other of the Milky Way of Magellanic Cloud laws, the difference in the
corrected flux would have been of order 10\% or less.  The resulting,
corrected monochromatic luminosities are also included in column (5)
of Table~\ref{T:indsed}. For reference, in column (4) of
Table~\ref{T:indsed} we also list the monochromatic luminosity after
correcting with the starburst extinction law of \citet*{calzetti94},
which yields the minimum correction compared to the other laws.

We draw attention to the uncertainties involved in our extinction
corrections: the most important uncertainty is the assumption of the
Milky Way dust-to-gas ratio, which is incorporated into
equation~(\ref{Q:extinction}), and our adopted value of $R_{\rm
V}=3.2$. We also emphasize that the extinction corrections we adopted
are suitable for our specific application (assessing the energy
budgets of LINERs; see \ptwo), but may not be appropriate for other
applications. More specifically, in \ptwo\ we estimate the ionizing
luminosities of the weak AGNs in this sample in order to assess
whether they can power the observed emission lines. Therefore, we take
the highest possible extinction correction so as to err on the side of
caution, i.e., to overestimate the UV luminosity rather than
underestimate it.

\def\fte{\tablenotemark{e}}
\def\ftf{\tablenotemark{f}}
\def\ftg{\tablenotemark{g}}
\def\od{\phantom{0}}
\def\td{\phantom{00}}
\begin{deluxetable*}{clrccrrrr}
\tabletypesize{\scriptsize}
\tablewidth{7in}
\tablecaption{X-Ray Spectral Parameters and Derived Properties\label{T:XUV}}
\tablehead{
\colhead{} &
\multicolumn{3}{c}{X-Ray Spectral Parameters} &
\colhead{} &
\colhead{} &
\colhead{} &
\colhead{} \\
\colhead{} &
\multicolumn{3}{c}{\hrulefill} &
\colhead{} &
\colhead{} &
\colhead{} &
\colhead{} \\
\colhead{Galaxy} &
\colhead{} &
\colhead{$N_0$} &
\colhead{$N_{\rm H}$} &
\colhead{$E(B-V)_{\rm X}$\tablenotemark{b}} &
\colhead{$L_{\rm 2-10\;keV}$} &
\colhead{$L_{\rm bol}$\tablenotemark{c}} &
\colhead{} &
\colhead{} \\
\colhead{(NGC)} &
\colhead{$\Gamma$\tablenotemark{a}} &
\colhead{(${\rm cm^{-2}~s^{-1}}$)} &
\colhead{(cm$^{-2}$)} &
\colhead{(mag)} &
\colhead{(\ergs)} &
\colhead{(\ergs)} &
\colhead{\REdd} &
\colhead{\aox\tablenotemark{d}}\\
\colhead{(1)} &
\colhead{(2)} &
\colhead{(3)} &
\colhead{(4)} &
\colhead{(5)} &
\colhead{(6)} &
\colhead{(7)} &
\colhead{(8)} &
\colhead{(9)}  
}
%
%
\startdata                                                                                                        
  0266 & 1.40     & $ 5.223\tten{-6}$ & 1.5 \tten{21} &  0.262 & $ 7.4\tten{40}$ & $ 2.2\tten{42}$ &   \dots       & ...     \\ 
  0404 & 2.50     & $ 4.490\tten{-6}$ & 2.3 \tten{20} &  0.040 & $ 3.9\tten{36}$ & $ 1.2\tten{38}$ & $ 4\tten{-6}$ & ...     \\ 
  1097 & 1.64     & $ 3.864\tten{-4}$ & 2.3 \tten{20} &  0.040 & $ 4.3\tten{40}$ & $ 8.5\tten{41}$ & $ 5\tten{-5}$ & 1.14    \\ %
  1553 & 1.20     & $ 4.438\tten{-5}$ & 3.2 \tten{21} &  0.559 & $ 8.7\tten{39}$ & $ 4.4\tten{41}$ & $ 4\tten{-5}$ & ...     \\ 
  2681 & 2.00     & $ 4.990\tten{-6}$ & 2.7 \tten{21} &  0.471 & $ 1.8\tten{38}$ & $ 9.0\tten{39}$ & $ 6\tten{-6}$ & ...     \\ 
  3031 & 1.88     & $ 1.775\tten{-3}$ & 4.1 \tten{20} &  0.072 & $ 1.9\tten{40}$ & $ 2.1\tten{41}$ & $ 3\tten{-5}$ & 1.04    \\ 
  3169 & 2.60     & $ 5.138\tten{-3}$ & 1.12\tten{23} &19.55\td& $ 1.1\tten{41}$ & $ 3.3\tten{42}$ & $ 4\tten{-4}$ & ...     \\ 
  3226 & 2.21     & $ 4.440\tten{-4}$ & 9.3 \tten{21} &  1.624 & $ 5.0\tten{40}$ & $ 1.5\tten{42}$ & $ 1\tten{-4}$ & ...     \\ 
  3379 & 1.80\fte & $ 6.200\tten{-7}$ & 2.75\tten{20} &  0.048 & $ 1.7\tten{37}$ & $ 5.1\tten{38}$ & $ 3\tten{-8}$ & ...     \\ 
  3507 & 1.80\ftf & $<2.400\tten{-7}$ & 1.63\tten{20} &  0.029 & $<3.9\tten{37}$ & $<1.2\tten{39}$ &   \dots       & ...     \\ 
  3607 & 1.80\ftf & $<3.000\tten{-7}$ & 1.48\tten{20} &  0.026 & $<5.0\tten{37}$ & $<1.5\tten{39}$ & $<5\tten{-8}$ & $>$0.55 \\ 
  3608 & 1.80\ftf & $<2.600\tten{-6}$ & 1.49\tten{20} &  0.026 & $<5.9\tten{38}$ & $<1.8\tten{40}$ & $<1\tten{-6}$ & ...     \\ 
  3628 & 1.80\ftf & $<2.000\tten{-7}$ & 2.23\tten{20} &  0.039 & $<4.9\tten{36}$ & $<1.5\tten{38}$ & $<2\tten{-8}$ & ...     \\ 
  3998 & 1.88     & $ 3.611\tten{-3}$ & 5.82\tten{20} &  0.102 & $ 2.6\tten{41}$ & $ 1.4\tten{43}$ & $ 4\tten{-4}$ & 1.05    \\ 
  4111 & 1.80\ftf & $<3.010\tten{-5}$ & 1.40\tten{20} &  0.024 & $<3.6\tten{39}$ & $<1.1\tten{41}$ & $<2\tten{-5}$ & $>$1.00 \\ 
  4125 & 1.80\ftf & $<2.200\tten{-6}$ & 1.84\tten{20} &  0.032 & $<5.4\tten{38}$ & $<1.6\tten{40}$ & $<6\tten{-7}$ & ...     \\ 
  4143 & 1.66     & $ 7.331\tten{-5}$ & 1.5 \tten{20} &  0.026 & $ 1.1\tten{40}$ & $ 3.2\tten{41}$ & $ 1\tten{-5}$ & ...     \\ 
  4261 & 1.56     & $ 1.900\tten{-4}$ & 5.0 \tten{22} & 8.73\od& $ 1.0\tten{41}$ & $ 6.8\tten{41}$ & $ 1\tten{-5}$ & ...     \\ 
  4278 & 1.64     & $ 1.800\tten{-4}$ & 3.5 \tten{20} &  0.061 & $ 9.1\tten{39}$ & $ 2.7\tten{41}$ & $ 5\tten{-6}$ & ...     \\ 
  4314 & 2.10     & $<1.130\tten{-4}$ & 1.78\tten{20} &  0.031 & $<3.1\tten{37}$ & $<9.2\tten{38}$ & $<5\tten{-7}$ & ...     \\ 
  4374 & 2.10     & $ 3.762\tten{-5}$ & 1.9 \tten{21} &  0.332 & $ 3.5\tten{39}$ & $ 5.0\tten{41}$ & $ 3\tten{-6}$ & 0.99    \\ 
  4438 & 1.80\ftf & $<9.999\tten{-6}$ & 1.20\tten{21} &  0.210 & $<1.2\tten{39}$ & $<3.5\tten{40}$ &  \dots        & $>$0.88 \\ 
  4457 & 1.70     & $ 6.815\tten{-6}$ & 9.8 \tten{20} &  0.171 & $ 1.0\tten{39}$ & $ 3.0\tten{40}$ & $ 3\tten{-5}$ & ...     \\ 
  4486 & 2.17     & $ 2.310\tten{-4}$ & 6.1 \tten{20} &  0.107 & $ 1.6\tten{40}$ & $ 9.8\tten{41}$ & $ 2\tten{-6}$ & 1.33    \\ 
  4494 & 1.80     & $ 2.352\tten{-5}$ & 3.0 \tten{20} &  0.052 & $ 9.2\tten{38}$ & $ 2.8\tten{40}$ & $ 6\tten{-6}$ & ...     \\ 
  4548 & 1.70     & $ 4.432\tten{-5}$ & 1.6 \tten{22} &  2.793 & $ 5.4\tten{39}$ & $ 1.6\tten{41}$ & $ 3\tten{-5}$ & ...     \\ 
  4552 & 2.00     & $ 6.700\tten{-6}$ & 6.0 \tten{20} &  0.105 & $ 2.6\tten{39}$ & $ 7.8\tten{40}$ & $ 4\tten{-6}$ & 1.03    \\ 
  4579 & 1.50     & $ 2.116\tten{-5}$ & 2.54\tten{20} &  0.044 & $ 1.8\tten{41}$ & $ 1.0\tten{42}$ & $ 1\tten{-4}$ & 0.92    \\ 
  4594 & 1.89     & $ 2.450\tten{-4}$ & 1.9 \tten{21} &  0.332 & $ 7.5\tten{39}$ & $ 4.8\tten{41}$ & $ 4\tten{-6}$ & 1.23    \\ 
  4636 & 1.80\ftf & $<5.100\tten{-6}$ & 1.81\tten{20} &  0.032 & $<6.1\tten{38}$ & $<1.8\tten{40}$ & $<1\tten{-6}$ & $>$1.14 \\ 
  4736 & 1.60     & $ 5.600\tten{-5}$ & 3.3 \tten{20} &  0.058 & $ 5.9\tten{38}$ & $ 1.8\tten{40}$ & $ 1\tten{-5}$ & 1.36    \\ 
  5055 & 1.80     & $ 2.843\tten{-5}$ & 5.0 \tten{21} &  0.873 & $ 2.0\tten{38}$ & $ 5.9\tten{39}$ & $ 4\tten{-6}$ & ...     \\ 
  5866 & 1.80\ftf & $<3.199\tten{-6}$ & 1.47\tten{20} &  0.024 & $<3.1\tten{38}$ & $<9.4\tten{39}$ & $<1\tten{-6}$ & $>$0.81 \\ 
  6500 & 3.10     & $ 5.178\tten{-5}$ & 2.1 \tten{21} &  0.367 & $ 5.3\tten{39}$ & $ 1.6\tten{41}$ & $ 7\tten{-6}$ & ...     \\ 
  7331 & 1.80\ftf & $<4.001\tten{-7}$ & 8.61\tten{20} &  0.150 & $<3.4\tten{37}$ & $<1.0\tten{39}$ & $<3\tten{-7}$ & $>$1.33 \\ 
\enddata
\tablenotetext{a}{The typical uncertainty in $\Gamma$ is $\pm 0.2$--0.3.
The fractional uncertainty in the X-ray flux is dominated by the
uncertainty in $\Gamma$ and is given by $\delta f_{\rm X} / f_{\rm X}
= 0.69 \;\delta\Gamma$. Thus, fractional uncertainties in the X-ray
flux are of order 10--20\%.}
\tablenotetext{b}{The reddening corresponding to the equivalent
hydrogen column density reported in this table, obtained using
equation~(\ref{Q:extinction}) in \S\ref{S:data} of the text.}
\tablenotetext{c}{The bolometric luminosity of NGC~3031,
NGC~3998, NGC~4261, NGC~4374, NGC~4486, NGC~4579, and NGC~4594 was
determined by integrating the S.E.D.  The bolometric luminosity of
NGC~1097 was derived from a model fit to the \sed\ (see
\S\ref{S:objects}). For all other galaxies the bolometric luminosity
was obtained by scaling the 2--10~keV X-ray luminosity as discussed
in \S\ref{S:aox} of the text.}
\tablenotetext{d}{The optical-to-X-ray spectral index, defined 
in \S\ref{S:aox} of the text. The tabulated values were computed after
correcting the UV flux according to the \citet{seaton79} law and using
the values of $E(B-V)_{\rm X}$ from column (5) of this table (maximal
correction). Typical uncertainties on \aox\ are $\gs 0.03$ and can be
as high as 0.09 (see the discussion in \S\ref{S:aox} of the text).}
\tablenotetext{e}{Because of the low S/N of the X-ray spectrum, 
The X-ray photon index was assumed to be $\Gamma=1.8$ for the 
purpose of deriving a value for the normalization. The normalization
itself and resulting X-ray flux are not upper limits, but 
are subject to the assumed value of $\Gamma$.}
\tablenotetext{f}{The X-ray photon index was assumed to be
$\Gamma=1.8$ for the purpose of deriving an upper limit to the
normalization and the X-ray flux.}
%
%
\end{deluxetable*}

\subsection{Notes on Individual Objects}\label{S:objects}

\begin{description}

\item[\it NGC 266. --] 
The radio properties of the AGN in NGC~266 have been studied by
\citet*{doi05a} using observations at multiple epochs.  They found
significant variability in both the radio luminosity and the shape of
the radio spectrum. Here we adopt their data from VLBA observations
taken on a single epoch, 2003 March 8.

\item[\it NGC 404. --] 
A UV spectrum of the nucleus of this galaxy shows prominent absorption
features from hot stars \citep{maoz98}, indicating that they make a
significant contribution to the light at these wavelengths. The X-ray
spectrum is very soft \citep[see the discussion in][]{eracleous02},
with $\Gamma=2.5$, which is uncharacteristic of AGNs
\citep[cf,][]{nandra97}, although an AGN cannot be ruled out based on
this observation.  The nuclear source is resolved at both UV and X-ray
wavelengths and it has a ``blow-out'' morphology
\citep{maoz95,eracleous02}. On the other hand, the compact nucleus of
the UV source appears to be variable by a factor of approximately two
on a time scale of approximately a decade. In addition, the depths of
the absorption lines appear to be shallower than the those found in
the spectra of hot stars \citep{maoz05,maoz07}. All of these
properties suggest that the nucleus of NGC~404 harbors a compact
star-forming region as well as a low-luminosity AGN.  Therefore, we
include this object in our sample.

\item[\it NGC 1097. --] 
The \sed\ of the AGN in NGC~1097 and model fits to it are discussed by
\citet{nemmen06}. Here, we adopted a subset of the data included in
that paper. The IR measurements presented by \citet*{prieto05} were
taken through a very small aperture, which isolates the
nucleus. However, the AGN is embedded in an compact, unresolved
starburst \citep{storchi05} which may dominate the emission at these
wavelengths, therefore, we have designated these measurements as upper
limits to the flux of the AGN.  We excluded measurements taken through
large apertures since these were significantly contaminated by
emission from the circumnuclear starburst ring. A significant fraction
of the nuclear UV flux appears to originate from a compact starburst,
as suggested by absorption lines from hot stars detected in the
\hst\ spectrum \citep{storchi05,nemmen06}. By following the
best-fitting \sed\ model of \citep{nemmen06}, we adopt only the
fraction of the UV flux that is attributed to the AGN. This model
includes contributions from an inner, radiatively inefficient
accretion flow, and outer, geometrically thin accretion disk, an
obscured starburst and a jet. The contribution of the jet is
appreciable only at the lowest radio frequencies, the starburst
contributes primarily to the near-UV band, the inner, hot accretion
flow dominates in the far-IR and X-ray bands, and the thin accretion
disk dominates in the near-IR band.  In the same spirit, we have also
adopted a bolometric luminosity of $8.5\times 10^{41}~\ergs$ from
\citet{nemmen06}. In comparison, if we integrate the tabulated SED, we
obtain a luminosity of $5.1\times 10^{41}~\ergs$.

\item[\it NGC 3998. --] 
To construct the \sed\ of NGC~3998 we began from the extensive data
tabulation of \citet{ptak04}. We excluded many of the measurements
presented therein because they were obtained through extremely large
apertures that encompass a substantial fraction of the host galaxy
(e.g., from {\it IRAS} observations). Since the AGN in NGC~3998 is
rather bright compared to other objects in our collection, we adopted
measurements through apertures as large as 3\asec\ as fair
measurements of the AGN luminosity. Measurements through apertures
between 3\asec\ and 15\asec\ were taken as upper limits to the AGN
luminosity.

\item[\it NGC 4261. --] 
The X-ray spectrum of the AGN in NGC~4261 has been measured recently
by both \chandra\ \citep{zezas05} and \xmm\ \citep{gliozzi03}. Both
observations yield the same spectral index and flux but significantly
different equivalent hydrogen column densities (the \chandra\ spectrum
yields $N_{\rm H}=3.7\times 10^{20}~{\rm cm}^{-2}$, while the \xmm\
spectrum yields $N_{\rm H}=5.0\times 10^{22}~{\rm cm}^{-2}$).  A UV
observation with the \hst, reported by \citet{zirbel98}, yields only
an upper limit of $\nLn < 4.9\times 10^{38}~\ergs$ at 2300~\AA, which
produces a very large dip in the \sed\ \citep[see Fig.~3 of][]{ho99}.
Neither of the column densities measured from the X-ray spectra
produces a reasonable extinction correction of the UV limit; the lower
value produces a negligible correction, while the higher value moves
the upper limit many orders of magnitude above the monochromatic X-ray
luminosity at 0.5~keV. Therefore, we infer that the AGN is indeed
completely extinguished in the UV by the large equivalent hydrogen
column measured from the \xmm\ spectrum. Thus we derive an upper limit
to the near-UV flux by assuming that $\aox < 1.5$, which we include in
Table~\ref{T:indsed}, and plot in Figure~\ref{F:indsed}.  Similarly,
we doubt that the measurements of the nucleus of NGC~4261 in the
optical band capture the AGN. Therefore, we do not apply any
extinction corrections to these measurements in Table~\ref{T:indsed}
nor to the points plotted in Figure~\ref{F:indsed}.

\item[\it NGC 5055. --] 
Prominent absorption lines in the UV spectrum of the nucleus of NGC
5055 suggest that hot stars dominate the light \citep{maoz98}. This
conclusion is supported by the lack of significant UV variability and
the extended morphology of the UV source \citep{maoz05}.  Although an
X-ray source with an AGN-like spectrum is detected \citep{flohic06},
the equivalent hydrogen column density measured from the X-ray
spectrum implies $E(B-V)=0.87$, which translates into an attenuation
by a factor of 100 at 2500~\AA.  Our conclusion is that an AGN may be
present in this galaxy, but is not the source of the observed UV flux.

\item[\it NGC 6500. --] 
The UV spectrum of the nucleus of NGC~6500 \citep{maoz98} has a
relatively low S/N, but still shows absorption features resembling
lines from hot stars. Moreover, the nuclear UV source is extended,
with no clear ``knot'' that could be identified with the nucleus
\citep{maoz95,barth98} and no variability \citep{maoz05}. The X-ray
spectrum \citep{terashima03} is indicative of an obscured AGN with
$E(B-V)=0.37$, implying an attenuation of its 2500~\AA\ flux by a
factor of 7. Thus this object is very similar to NGC~5055; an AGN is
probably present, but is not the source of the observed UV flux.

\end{description}

\section{Quantities Derived from the Spectral Energy Distributions}\label{S:derived}

\subsection{Optical-to-X-Ray Spectral Indices, Bolometric Luminosities, and Eddington Ratios}\label{S:aox}

For comparison with other types of AGNs, we have used the rest-frame
flux densities at 2500~\AA\ and 2~keV to compute the
``optical-to-X-ray spectral index'' \citep{tananbaum79}, defined as
$$
\aox \equiv 
-{\log\left[L_{\nu}(2500\;{\rm \AA})/L_{\nu}(2~{\rm keV})\right]
\over 
\log\left[\nu(2500\;{\rm \AA})/\nu(2~{\rm keV})\right]}
$$
\begin{equation}
= 1+0.384\; \log\left[
{(\nu L_{\nu})_{\rm 2500\;\AA}\over(\nu L_{\nu})_{\rm 2\; keV}}
\right]\; ,
\label{Q:aox}
\end{equation}
under the convention that $L_{\nu}\propto\nu^{-\aox}$. Both the UV and
X-ray flux densities were corrected for extinction, as described in
\S\ref{S:extinction}, so that \aox\ describes the shape of the
intrinsic AGN spectral energy distribution. We were able to calculate
the values of \aox\ for the 9 LINERs with data available at
2500~\AA\ and 2~keV.  Another 6 objects have available data at
2500~\AA\ but only upper limits at 2~keV; for these we were able to
obtain lower limits to \aox. For three objects, NGC~404, NGC~6500, and
NGC~5055, we do not present the value of \aox\ because we feel it is
unreliable (see the discussion of these three objects in
\S\ref{S:objects}). In the case of NGC~404, although an AGN may
contribute to observed UV and X-ray flux, there is significant
contamination by hot stars, as evidenced by stellar absorption lines
in the UV spectrum. In the other two objects, the UV light that we
observe appears to be dominated by hot stars, with no discernible
contribution from an AGN.

The uncertainty in \aox\ is related to the fractional uncertainties in
the X-ray and UV fluxes (as long as these are small) via $\delta\aox =
0.17\left[(\delta f_{\rm X}/f_{\rm X})^2+(\delta f_{\rm UV}/f_{\rm
    UV})^2\right]^{1/2}$. Typically, $\delta f_{\rm X}/f_{\rm
  X}\approx 0.1$--0.2 [dominated by uncertainties in the photon index;
  see footnote (a) in Table~\ref{T:XUV}]\footnote{The uncertainty on
  $f_{\rm X}$ should also have contribution from variability, which we
  are unable to estimate because of the lack of systematic monitoring
  data.}. The value of $\delta f_{\rm UV}/f_{\rm UV}$ includes a
contribution from variability (typically, $\delta f_{\rm UV}/f_{\rm
  UV} \sim 0.10$ for small-amplitude variability on time scales of
1--2 years; see Table~\ref{T:indsed}) and a contribution from
uncertain extinction corrections (the difference between the starburst
and Milky Way or Magellanic Cloud extinction laws leads to $\delta
f_{\rm UV}/f_{\rm UV} < 0.13$ for 2/3 of our objects). Thus, under the
assumption of small errors, the uncertainty in \aox\ is 0.03. However,
the above analysis does not apply if the amplitude of variability is
large \citep[it could reach a factor of a few over the course of
  several years; see][]{maoz05} or if $E(B-V)_{\rm X} > 0.2$ and the
extinction law is uncertain. In such a case, the magnitude of the
uncertainty is best illustrated by the following specific examples. If
the X-ray or UV flux change by a factor of 3 between the observations
in the two bands, then \aox\ will change by 0.2. For an object such as
NGC~4438 with $E(B-V)_{\rm X} = 0.21$ the maximal difference between
extinction laws yields a change in \aox\ of 0.24, while for an object
such as NGC~2681 with $E(B-V)_{\rm X} = 0.47$, the change in
\aox\ is 0.50.

\begin{figure}
\centerline{\includegraphics[scale=0.45,angle=0]{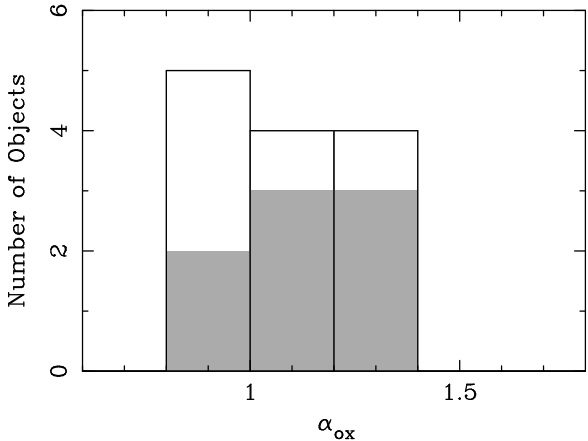}} 
\caption{ Distribution of the values of the optical-to-X-ray spectral
index, \aox, for 19 of the galaxies in our sample.  The values were
computed via equation~(\ref{Q:aox}). The hollow bins represent lower
limits, which correspond to cases where the 2500~\AA\ UV flux has been
measured but only an upper limit to 2~keV X-ray flux is available. 
\S\ref{S:objects}.
\label{F:aox}}
\end{figure}

The resulting values of \aox\ are given in Table~\ref{T:XUV} and their
distribution is plotted in Figure~\ref{F:aox}.  The majority of the
values of \aox\ (excluding limits) cluster between 0.92 and 1.36,
while the lower limits fall between 0.55 and 1.33.  It is useful to
compare the values that we obtain here with those obtained by
\cite{maoz07} using similar data and methodology. In their sample of
12 objects (excluding NGC~404) \citep{maoz07} find a range of \aox\ of
0.92--1.34, very similar to ours. Moreover, comparing the values of
\aox\ in five individual objects that are common between the two
papers, we find that in four cases they agree within 0.16 and in the
case of NGC~4594 the values disagree by 0.31. We attribute the
differences in the values of \aox\ to the following differences in the
analysis: (a) \citet{maoz07} did not apply corrections for intrinsic
extinction in the vicinity of the AGN, while we did, and (b)
\citet{maoz07} used a fixed value of 1.8 for the X-ray photon index to
obtain the 2~keV monochromatic luminosity, while we used the specific
value of the photon index measured from each X-ray spectrum. Thus we
conclude that the true range of values is likely between 0.9--1.4. We
take 1.5 to be a very generous upper limit to the value of \aox\ and
use it to derive upper limits to the 2500~\AA\ monochromatic
luminosity of the objects for which no such measurement or limit is
available (see \S\ref{S:data}).

We have computed the bolometric luminosities of a set of ``calibration
objects'' with well-sampled \seds\ by integrating the \seds\ directly
(neglecting upper limits). These objects are NGC~3031, NGC~3998,
NGC~4374, NGC~4486, NGC~4579, and NGC~4594. For another object,
NGC~1097, we adopted the bolometric luminosity derived from a model
for the \sed\ by \citet{nemmen06}. To integrate an {\sed}s we assumed
that pairs of neighboring points defined a power law, computed the
integral for each segment analytically, and summed the luminosities of
individual segments. Using this small subset of objects, we estimated
the ``bolometric correction,'' \zbol, such that $L_{\rm bol} = \zbol\;
L_{\rm 2-10\; keV}$. We found that the values of \zbol\ in this small
subset span a wide range, from 6 to 143, with a geometric mean of 33,
a median of 52, and an average of 51. This range is comparable to the
range found in Seyfert galaxies and quasars by \citet{vasudevan07}.
Thus, we adopted $\zbol=50$ and used it to obtain the bolometric
luminosity of the remaining objects in our sample from their 2--10~keV
X-ray luminosities. We list these bolometric luminosities in
Table~\ref{T:XUV}.

\begin{figure}
\centerline{\includegraphics[scale=0.45,angle=0]{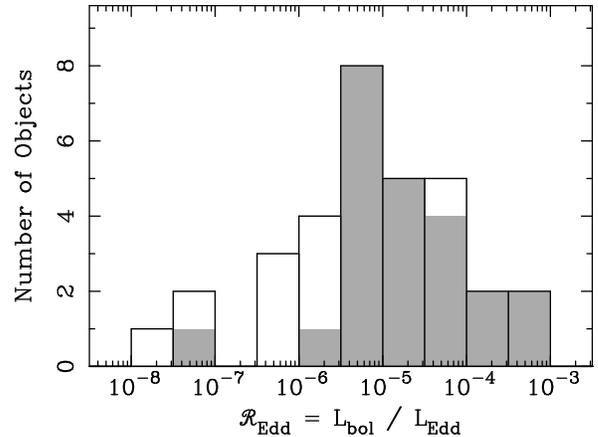}} 
\caption{Distribution Eddington ratios for the AGNs in our
sample. These were computed via equation~(\ref{Q:edd}), with
bolometric luminosities estimated as described in \S\ref{S:sed} of the
text. The hollow histogram bins represent upper limits. The galaxies
NGC~266, NGC~3507, and NGC~4438 are not included because their black
hole masses are not available.
\label{F:REdd}}
\end{figure}

Using the bolometric luminosities estimated above, we computed the
Eddington ratios of the AGNs in our sample (the ratio of the bolometric
luminosity of the AGN and the Eddington luminosity) as follows:
$$
\REdd\equiv{L_{\rm bol}/L_{\rm Edd}} = 7.7\times 10^{-7}\; L_{40}\; M_8^{-1} 
$$
\begin{equation} 
{\rm or} \quad 
\log \REdd = -6.11 + \log L_{40} - \log M_8\; , 
\label{Q:edd}
\end{equation}
where $L_{\rm bol} = 10^{40}\; L_{40}~{\rm erg~s}^{-1}$ is the
bolometric luminosity, $L_{\rm Edd} = 1.3\times 10^{38}\; (M_{\rm
BH}/\Msol)~{\rm erg~s}^{-1}$, is the Eddington luminosity, and $M_{\rm
BH}=10^8 M_8~\Msol$ is the black hole mass reported in Table~1.  The
resulting values of $\REdd$ are included in Table~\ref{T:XUV}; they
span the range $-7.8 < \log \REdd < -3.4$ for the AGNs in our sample
that were {\it detected} in the X-ray band. The distribution of
Eddington ratios is shown graphically in Figure~\ref{F:REdd}.

Estimating the bolometric luminosity by scaling the 2--10~keV
luminosity is subject to a number of caveats. These caveats also apply
to the estimates of the Eddington ratio. First, it is not known
whether there is a single, universal value of the bolometric
correction. On theoretical grounds, we would expect the shapes of the
\seds\ of low-Eddington ratio AGNs, hence the bolometric corrections,
to depend on the black hole mass and accretion rate \citep[see, for
example,][and references therein]{ball01}. The objects in our
calibration set span a relatively narrow range in luminosity, mass,
and Eddington ratio [$41.3 < \log (L_{\rm bol}/\ergs) < 43.1$, $7.8 <
\log (\Mbh/\Msol) < 9.5$, $-3.8 < \log \REdd < -5.3$], but we have
applied this correction to objects with a much wider range in these
properties, namely down to $\log (\Mbh/\Msol) = 5.3$ and $\log \REdd =
-7.8$. We have checked whether there is any correlation between \zbol\
or \aox, and the mass, luminosity, or Eddington ratio in our
calibration set but found none. Second, the values of \zbol\ within
our calibration set span a rather wide range even though they were
derived from objects with a narrow range of properties. This suggests
that either (a) there are systematic errors in the measured bolometric
luminosities, or (b) there is a wide diversity in the
\seds. Systematic errors could arise from our interpolating over broad
but sparsely-sampled frequency windows (especially in the far-IR) or
from contamination by light from the host galaxy (especially in the
near-IR). In conclusion, we note that the values of the bolometric
luminosity and the Eddington ratio that we report in Table~\ref{T:XUV}
should be regarded with caution; we estimate rough uncertainties on
these quantities of a factor of 5 in either direction.

\begin{figure}
\centerline{\includegraphics[scale=0.5,angle=0]{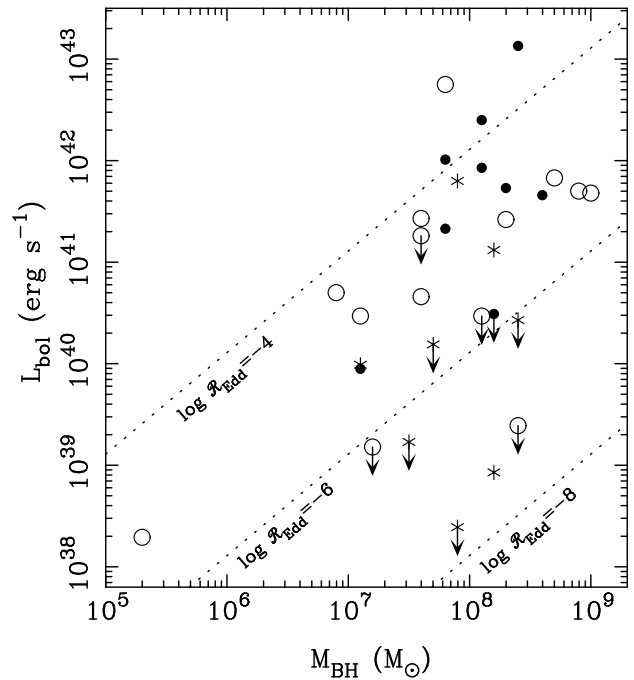}} 
\caption{Bolometric luminosities (see \S\ref{S:aox}) plotted against
black hole mass.  The galaxies NGC~266, NGC~3507, and NGC~4438 are not
included because their black hole masses are not available. Arrows
denote upper limits on the bolometric luminosity, while different
symbols represent different LINER classes: L1 objects are plotted as
filled circles, L2 objects as open circles, and T2 and two L2/T2
objects as asterisks. For reference, diagonal dotted lines indicate
the locus of constant Eddington ratio for $\log\REdd=-4,\;-6,\;-8$
(from top to bottom).
\label{F:Lbol}}
\end{figure}

In an effort to assess any relation between LINER class and AGN
luminosity, which may reveal selection effects or other biases in this
sample, we plot in Figure~\ref{F:Lbol} the bolometric luminosities we
have derived against the black hole mass. In this figure we use
different symbols to denote different LINER classes (L1, L2, and T2 or
T2/L2). An inspection of this figure shows that L2 and T2 LINERs span
the entire range of luminosities of the sample, although L2 objects
are concentrated at higher luminosities. This is not a surprise
because more luminous objects, which also appear brighter in the
narrow range of distances of this sample, are more attractive targets
for pointed X-ray observations. L1 LINERs are found preferentially at
high luminosities, which we attribute to a selection effect of the
optical spectroscopic survey that detected them \citep{ho97c}. More
specifically, the luminosity of their broad lines, which determines
how easily they can be detected, depends directly on the luminosity of
the AGN.

\begin{figure*}
\centerline{\includegraphics[scale=0.4,angle=-90]{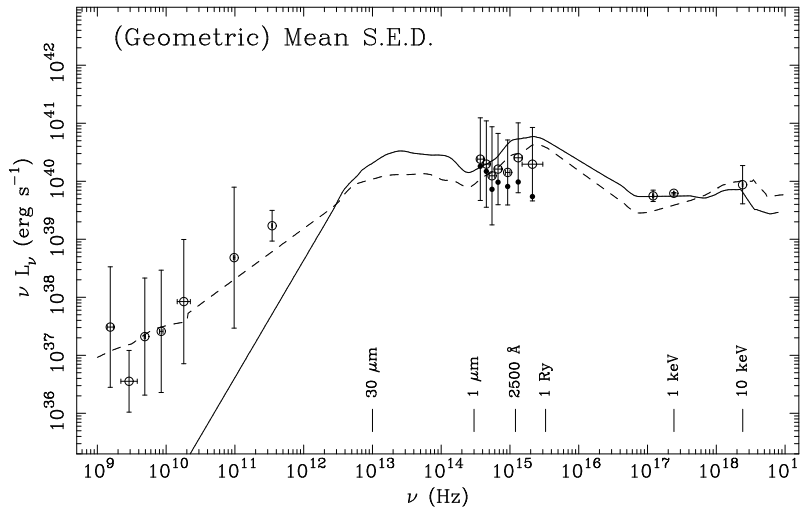}} 
\caption{The geometric mean \sed\ of the LINERs in our sample.  The
open circles represent the average after extinction corrections, while
the filled circles show the average before any extinction corrections
are applied (in the range 0.1--1~$\mu$m only).  The vertical error
bars indicate the standard deviation in $\log(\nu L_{\nu})$.  The
horizontal error bars represent the frequency bands within which
measurements were averaged. Overplotted are the average \seds\ of
radio-quiet and radio-loud quasars \citep[solid and dashed lines
respectively; from][]{elvis94}. Additional details can be found in
\S\ref{S:avgsed} of the text and in Table~\ref{T:avgsed}.
\label{F:avgsed}}
\end{figure*}

\begin{deluxetable*}{llccccc}
\tabletypesize{\scriptsize}
\tablewidth{5in}
\tablecaption{Average Spectral Energy Distribution\tablenotemark{a}\label{T:avgsed}}
\tablehead{
\multicolumn{2}{c}{Band Limits\tablenotemark{b}} &
\colhead{} &
\colhead{} &
\multicolumn{2}{c}{Corrected\tablenotemark{d}} &
\colhead{} \\
\multicolumn{2}{c}{\hrulefill} &
\colhead{Number} &
\colhead{} &
\multicolumn{2}{c}{\hrulefill} &
\colhead{Uncorrected\tablenotemark{e}} \\
\colhead{$\nu_{\rm min}$} &
\colhead{$\nu_{\rm max}$} &
\colhead{of Points\tablenotemark{c}} &
\colhead{$\log\langle\nu\rangle$} &
\colhead{$\langle\log (\nu L_{\nu})\rangle$} &
\colhead{$\sigma\left[\log(\nu L_{\nu})\right]$\tablenotemark{f}} &
\colhead{$\langle\log (\nu L_{\nu})\rangle$} \\
\colhead{(1)} &
\colhead{(2)} &
\colhead{(3)} &
\colhead{(4)} &
\colhead{(5)} &
\colhead{(6)} &
\colhead{(7)}  
}
\startdata                                                                       
 1.390\tten{ 9} & 1.710\tten{ 9} & 10 &  9.19 & 37.49 & 1.04 &  \\
 2.200\tten{ 9} & 3.800\tten{ 9} &  4 &  9.48 & 36.55 & 0.53 &  \\
 4.800\tten{ 9} & 5.000\tten{ 9} & 14 &  9.69 & 37.32 & 1.01 &  \\
 8.000\tten{ 9} & 9.000\tten{ 9} &  7 &  9.93 & 37.41 & 1.05 &  \\
 1.450\tten{10} & 2.250\tten{10} & 15 & 10.27 & 37.93 & 1.07 &  \\
 9.500\tten{10} & 1.000\tten{11} & 10 & 10.99 & 38.68 & 1.21 &  \\
 3.450\tten{11} & 3.500\tten{11} &  3 & 11.54 & 39.23 & 0.26 &  \\
 3.333\tten{14} & 4.110\tten{14} & 10 & 14.57 & 40.38 & 0.71 & 40.26 \\
 4.110\tten{14} & 5.000\tten{14} &  5 & 14.66 & 40.30 & 0.74 & 40.17 \\
 5.000\tten{14} & 6.000\tten{14} &  8 & 14.74 & 40.09 & 0.85 & 39.86 \\
 6.000\tten{14} & 7.500\tten{14} &  4 & 14.83 & 40.21 & 0.62 & 39.98 \\
 8.571\tten{14} & 1.000\tten{15} &  8 & 14.97 & 40.15 & 0.56 & 39.91 \\
 1.154\tten{15} & 1.500\tten{15} & 14 & 15.12 & 40.40 & 0.60 & 39.99 \\
 1.500\tten{15} & 3.000\tten{15} &  8 & 15.35 & 40.29 & 0.63 & 39.74 \\
 1.182\tten{17} & 1.231\tten{17} & 21 & 17.08 & 39.75 & 0.10 &  \\
 2.389\tten{17} & 2.437\tten{17} & 21 & 17.38 & 39.79 & 0.01 &  \\
 2.389\tten{18} & 2.437\tten{18} & 21 & 18.38 & 39.94 & 0.33 &  \\
\enddata
\tablenotetext{a}{Frequencies are measured in Hz and monochromatic
luminosities ($\nu L_{\nu}$) are measured in \ergs.}
\tablenotetext{b}{The limits of the frequency bands used to group 
the data points from individual \seds.}
\tablenotetext{c}{The number of data points in each frequency bin.}
\tablenotetext{d}{Average values of $\log(\nu L_{\nu})$ and its
standard deviation {\it after} correcting for extinction using the
\citet{seaton79} law.}
\tablenotetext{e}{Average values of $\log(\nu L_{\nu})$ from 0.1 to 1~$\mu$m
{\it without} extinction corrections.}
\tablenotetext{f}{The standard deviation in the {\it corrected} values
of $\log(\nu L_{\nu})$ (omitted when a bin contains only a single data
point).}
\end{deluxetable*}

\subsection{The Average Spectral Energy Distribution and Its Properties}\label{S:avgsed}

For the sake of completeness, we have also produced an average
\sed\ using the following procedure. First, all individual \seds\ were
normalized (arbitrarily) to an integrated 2--10~keV X-ray luminosity
of $10^{40}~\ergs$. Because of this normalization scheme, we had to
exclude 11 AGNs for which we only have X-ray upper limits.  After
normalization, the available measurements were grouped into frequency
bins, neglecting upper limits as well as the \seds\ of NGC~404,
NGC~5055, and NGC~6500, which appear to be significantly contaminated
or dominated by starlight (see \S\ref{S:objects}). Finally, we
computed the geometric mean in each bin by averaging the logarithm of
the luminosity, $\langle\log(\nu L_{\nu})\rangle$. The geometric mean
is preferable to the simple mean because it minimizes the effect of
extreme outliers.  We have found that the geometric mean is very
similar to the median in each frequency bin. For frequency bins
between 0.1 and 1~$\mu$m we have computed the geometric mean both
before and after applying extinction corrections.  The results are
tabulated in Table~\ref{T:avgsed}, which also includes the definition
of the frequency bins used, the standard deviation of $\log(\nu
L_{\nu})$, and the number of measurements in each frequency bin. The
X-ray photon index in the (geometric) mean \sed\ is 1.85 and the value
of \aox\ is 1.24. In comparison, the subset of objects with a measured
value of \aox\ have a median X-ray photon index of 1.90 and a median
\aox\ of 1.28.

The mean \sed\ is plotted in Figure~\ref{F:avgsed} where we also
overplot the mean \seds\ of radio-loud and radio-quiet quasars
\citep{elvis94}, after normalizing them to the same X-ray luminosity,
for comparison. It is noteworthy that the mean LINER \sed\ has an
X-ray band slope similar to that of radio-quiet quasars but a relative
radio luminosity and slope similar to that of radio-loud quasars. In
the UV band, around 1~Ry, the luminosity in the quasar SEDs is
somewhat higher than in the mean LINER \sed, although there is
agreement within the 1-$\sigma$ dispersion limits.  It is also
interesting to compare the mean LINER \sed\ derived here with the
LINER \seds\ presented by \citet[][see his Fig.~8]{ho99}. The near-UV
luminosity in our \sed\ appears to be higher (relative to the X-ray
luminosity) than that of the \citet{ho99} \seds. This is also apparent
from a comparison of the values of \aox\ determined by \citet{ho99}
and by us.  We attribute this difference to newer measurements in both
the UV and X-ray bands. Measurements in the X-ray band with \chandra\
have isolated the X-ray emission from the AGN and reduced the
contamination from circumnuclear sources \citep[e.g.,][]{flohic06},
thus increasing the contrast between the UV and X-ray luminosity.

We emphasize that interpretation or use of the average \sed\ requires
care for the following reasons. First and foremost, the average
\sed\ includes contributions from objects with a wide range of
Eddington ratios, with the caveats discussed at the end of
\S\ref{S:aox}. We do note, however, that the dominant contribution is
from objects with $-5.4 < \log\REdd < -4.0$ (see
Fig.~\ref{F:REdd}). Second, not all objects contribute to all
frequency bins of the average \sed.  Each bin includes a different mix
of objects, with the exception of the X-ray bins, where by
construction all objects contribute. Third, in frequency bins
containing a small number of measurements, the standard deviation is
depressed.  In spite of these caveats, however, one can use more
specific filters for the data tabulated here in order to construct an
average \sed\ that is suitable for testing accretion flow models.

\acknowledgements 

We thank the referee, D. Maoz for many helpful comments and
E. C. Moran for a careful and critical reading of the manuscript.
This work was partially supported by the National Aeronautics and
Space Administration through {\it Chandra} award number AR4-5010A
issued by the {\it Chandra} X-Ray Observatory Center, which is
operated by the Smithsonian Astrophysical Observatory for and on
behalf of the National Aeronautics and Space Administration under
contract NAS8-03060. This research has made use of the NASA/IPAC
Extragalactic Database (NED) which is operated by the Jet Propulsion
Laboratory, California Institute of Technology, under contract with
the National Aeronautics and Space Administration.


\end{document}